# Electron spin dynamics of two-dimensional layered materials

*Bálint Náfrádi*[*], *Mohammad Choucair*[*], *László Forró*

Dr. B. Náfrádi, Prof. L. Forró

Institute of Condensed Matter Physics, École Polytechnique Fédérale de Lausanne, Lausanne 1015, Switzerland.
E-mail: balint.nafradi@epfl.ch

Dr. M. Choucair
School of Chemistry, University of Sydney, Sydney 2006, Australia.
E-mail: mohammad.choucair@sydney.edu.au

**Abstract**

The growing library of two-dimensional layered materials is providing researchers with a wealth of opportunity to explore and tune physical phenomena at the nanoscale. Here, we review the experimental and theoretical state-of-art concerning the electron spin dynamics in graphene, silicene, phosphorene, transition metal dichalcogenides, covalent heterostructures of organic molecules and topological materials. The spin transport, chemical and defect induced magnetic moments, and the effect of spin-orbit coupling and spin relaxation, are also discussed in relation to the field of spintronics.

# 1. Introduction

Spin electronics (spintronics) can be understood simply by assuming that any current of spins is carried by discrete spin-up and spin-down quantum states. This two-bit scheme of spin transport was used as the basis of solid-state spintronic devices that have been successfully employed commercially, like non-volatile magnetic random access memory (MRAM), that utilises magnetoresistance[1], and also in the development of the spin valve – working on the same principle[2]. Studies of the spin dynamics of these first-generation spintronic devices were not central to the phenomena probed[3]. However, there has been a growing emphasis on understanding spin dynamics in materials for the development of the second generation of spintronic devices exploiting spin coherence phenomena. These spin coherent properties include the precession of the magnetisation in nanomaterials[4, 5] and the spin Hall effect[6].

There are advantages of using spin-only based devices over that of electron-charge-based architectures[7]. For example, operations involving switching from one information state to another (e.g. 0 and 1) using electron spin would not need to raise or lower a barrier to charge motion if these operations are done coherently by application of a magnetic field (described below)[8]. The thermodynamic limitation on minimum switching energy that applies to charge-based switching devices[9] ($E = kT\ln2 \sim 23$ meV) does not apply to spin-based devices. By remaining out of equilibrium for periods of time of the order of the spin coherence times, this could, in principle, allow for more efficient spintronic devices that would be able to perform multiple independent operations before the carriers reach thermal equilibrium. If the spin transport phenomena (underlying present day technologies like MRAM) could be operational at room temperature in easily spin-polarisable materials, it would be possible to integrate non-volatile storage directly into logical processors. Hence, studies of the spin precession are critical to monitor and control the relative phases of the spin-up and spin-down components of the electron wave function to ensure robust operation in spintronics devices.

The most well-known approach used to coherently drive transitions between the Zeeman-split levels of an electron is electron spin resonance (ESR), whereby an oscillating magnetic field, $B_1$, is applied perpendicularly to the static field $B_0$ resonating with the spin precession frequency $f = g\mu_B B_0/h$ ($\mu_B$ is the Bohr magneton and $g$ is the electron spin $g$-factor, $h$ is the Planck's constant)[10] (Figure 1). In practice, this requires excitation in the microwave regime, as $B_0$ must be well above the geomagnetic field strength[11]. Using ESR, the spin precession of an ensemble of spins is used to infer information about the decoherence of individual electron spins. A detailed review on spins in few-electron materials, dubbed 'single spintronics', is available elsewhere[12].

For electron spins in 2-dimensional layered materials the most important interactions with the environment occur via the spin-orbit coupling and the hyperfine coupling with the nuclear spins of the host material. The spin-orbit coupling can lead to spin relaxation via several mechanisms like the D'yakonov-Perel' [13] and the Elliot-Yafet [14]. A detailed account of the mechanisms of electron spin relaxation have been addressed in detail elsewhere[12, 15]. Briefly, the energy relaxation processes of electron spins are described by a time constant $T_1$ which unavoidably also leads to the loss of quantum coherence described by a time constant $T_2$. Spin-phonon coupling occurs mostly indirectly, mediated either by the hyperfine interaction or by spin-orbit interaction, and, as a result, determine the value of $T_1$. If the effect of the nuclear field on the electron spin coherence could be suppressed, the spin-orbit interaction would limit $T_2$ to a value of $2T_1$. In magnetically homogenous itinerant systems (*e.g.* metals), the condition $T_1=T_2$ is often met and represents the longest period of time that in-phase precessing electron spins and magnetisation can propagate as a uniform mode[16]. The effect of the spin-orbit and hyperfine interactions can be observed in several ways including a deviation of the *g*-factor of electrons from the value of 2, and fluctuations in the magnetic fields in the environment leading to phase randomisation, respectively.

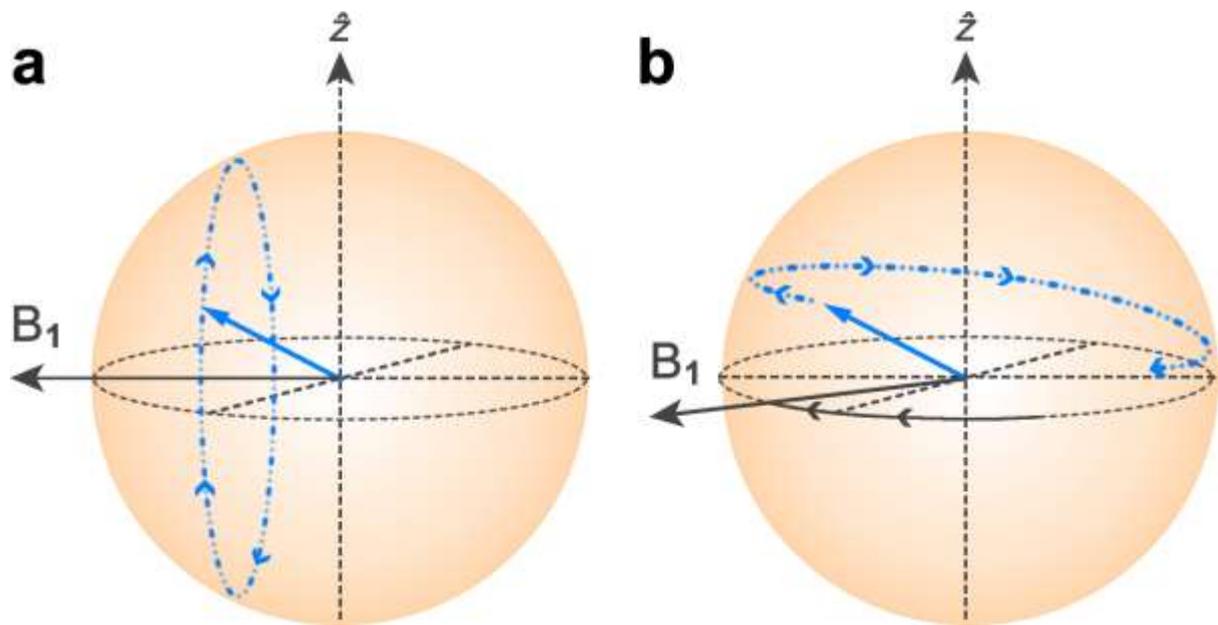

**Figure 1.** Motion of the electron spin during a spin-resonance experiment. (a) The motion as seen in a reference frame that rotates about the $\hat{z}$-axis at the same frequency *f* as the spin itself and the resonant rotating magnetic field $B_1$. The rotating field $B_1$ lies along a fixed axis in this rotating reference frame. An observer in the rotating frame will see the spin precession about $B_1$. (b) An observer from an external reference frame sees the spin spiral down over the surface of the Bloch sphere (a geometrical representation of a two-level quantum system)[8].

Foreign atoms in two-dimensional materials, e.g., adatoms, substituted atoms, and vacancies, can carry a magnetic moment[17-20]. Doping of two-dimensional materials with magnetic atoms can produce changes in the electronic structure of the material[19]. Dopants may cause impurity bands in the band gap region of the doped material that are almost completely spin-polarised. In the absence of *d* or *f* electrons, magnetism in *s* or *p* electron two-dimensional materials may exist under a variety of structural and chemical modifications that generate unpaired electrons[18, 20, 21]. The presence of conduction electrons provides a medium for magnetic coupling (antiferromagnetism and ferromagnetism) between localised spins and may also contribute to electron transport[22, 23]. However, achieving such magnetic alignment between unpaired electron spins has proven to be difficult[24]. It is likely that magnetic defects are responsible for short relaxation times in two-dimensional materials and contribute to completely quenching spin relaxation by doping above a certain threshold[25, 26]. The atomic doping of two-dimensional materials remains one of the major approaches to alter the magnetic properties in intrinsically non-magnetic two-dimensional materials.

From the principles outlined above we necessarily arrive at formulating a criterion for materials being considered for the next generation of room-temperature spintronic applications. An essential requirement, and indeed the prerequisite, is an intrinsic long carrier spin lifetime, with the shorter of $T_1$ and $T_2$ determining this value. For practical applications, this needs to exceed 100 ns, as this is currently the state of the art lower bound for signal processing times in quantum electronic devices[27]. Requirements of 2-dimensional materials for spintronics are numerous and arguable. For instance, in the context of spin transport, carrier mobility should be high – as when it is taken together with the spin lifetime – it determines the spin diffusion length[11, 28-30]. The combination of the values of these physical properties ultimately determines the feasibility of having practical device dimensions. In this regard 2-dimensional materials are desirable because their geometry is directly compatible with the established device designs and processing technologies already used in the semiconductor industry. The spin system itself needs also to be prudently considered: conduction electron spins provide more robustness against decoherence compared to localised spin states; the effects of internal fluctuating magnetic fields are dominant in localised spin systems which are minimised or absent in conducting or metallic materials as motional narrowing processes cancel these effect[23]. Other highly desirable qualities of the materials employed include abundance, easy and reproducible preparation, chemical and thermal stability, non-toxicity, and well-defined size and quantum characteristics.

Among the 2-dimensional crystals, graphene is one of the most promising material[31, 32], for spintronics due to its micrometre-scale ballistic transport at room temperature[28], to high carrier mobilities in suspended devices (~10000 cm$^2$ V$^{-1}$ s$^{-1}$ to ~200000 cm$^2$ V$^{-1}$ s$^{-1}$ at room-temperature)[33], and to spin lifetimes reaching 50 ns at room temperature (ref. [23, 34]). The search for new elemental 2-dimensional layered materials has intensified,[35, 36, 37] which has resulted the synthesis of phosphorene and silicene, from phosphorus and silicon, respectively. The potential for silicene in spintronics derives from the possibility of a controllable spin Hall effect[38, 39], and more generally, the emergence of coupled spin and valley degrees of freedom to route spins for logic operations.[40] Both silicene and phosphorene offer the possibility of facile methods to tune their band gaps[41, 42] that is desirable in nanoelectronics. Phosphorene-layered materials have demonstrated rapidly improving room-temperature mobilities[29, 43, 44] (reaching ~1350 cm$^2$ V$^{-1}$ s$^{-1}$) that are promising for spintronic applications, together with a tunable band gap that lies nicely between the zero band gap graphene and large band gap transition metal dichalcogenides (TMDs)[29, 42, 43].

Although single- and few-layer examples of TMDs have been synthesised[37], like MoS$_2$, MoSe$_2$, WS$_2$, and ZrS$_2$, it was predicted using first-principles density functional theory (DFT) for structure optimisation and phonon calculations that not all of the possible 88 different combinations of a transition metal (*M*) with a dichalcogenide (*X*) (i.e. *MX*$_2$) compounds can be stable in free-standing, single-layer honeycomb-like structures – rather, 44 stable monolayer *MX*$_2$ sheets have been nicely summarised and reported.[45] Single-layer *MX*$_2$ sheets can be semiconductors, ferromagnets, or nonmagnetic metals. These TMDs possess direct band gaps (1-2 eV) that are different not only from other 2-dimensional materials like graphene, but also from their bilayer, few-layer, and bulk phases. It is worth noting that the major difference between TMDs in relation to electron spin dynamics is the stronger spin-orbit coupling in the tungsten (W) compounds.

The structural and chemical rigidity in inorganic 2-dimensional materials is now being challenged through the judicious fabrication of stacked organic 2-dimensional materials[35, 46]. Due to the strong anisotropy of the layered components, even covalently bound layers could exhibit decoupled electronic properties. These covalently bound compounds – mostly molecular metals and charge-transfer salts – provide an unprecedented opportunity for finer control of the magnetic properties of spintronic materials down to the molecular level in a sort of a 'self-assembly' manner. The opportunity for these organic two-dimensional materials to be used in spintronics becomes very promising with the emergence of a new class of multiferroic materials like κ-(BEDT-TTF)$_2$Cu[N(CN)$_2$]Cl offering long spin diffusion lengths (0.2 μm) and spin lifetimes conduction electrons on the order of a few nanoseconds at room temperature[47, 48].

One of the latest developments which has not been yet explored in spintronics context are the metallic states on material surfaces which have emerged from non-trivial topological order[49]. These surface states of topological materials exhibit strong spin orbit coupling interactions and are forbidden to undergo back scattering. This symmetry protection provides a robustness against spin decoherence and has paved the way to non-trivial phases of matter. With the emergence of these new quantum states of matter, which include both topological insulators[50, 51] and Weyl semimetals[52-54], several device concepts are now taking advantage of topologically protected states. Experimental demonstrations of these concepts are in their infancy, although hold promise due to the potential ease at which stoichiometric control of bulk crystals can now be achieved.

In order to keep this feature article concise, we restrict ourselves to the reports relating to the electron spin dynamics of the elemental 2-dimensional materials including graphene, phosphorene, silicene, some TMDs namely MoS$_2$, MoSe$_2$, WS$_2$, ZrS$_2$, and an emerging class of 2-dimensional heterostructures made by the covalent stacking of molecular crystals on top of each other unlike the Van der Waals conceptual counterpart[35], and topological materials split into topological insulators and Weyl semimetals. There are reviews that address the structural and chemical properties of these nanostructures and those which are similar and provide perspectives other than our own[35, 36, 37, 55-58]. We aim to cover the latest understanding of the quantum spin properties and factors that greatly influence the adoption of these 2-dimensional layered materials in spintronic devices. We discuss with some emphasis the methods of fabricating the nominated 2-dimensional layered materials, the challenges of integrating them into working devices, and the relationship between their spin-electronic properties and stability. Materials with a considerable aspect of their research that remains theoretical, or those materials which systematically display similar properties only differing in the nature of constituent atoms, or where little is known about their electron spin properties, lie outside the scope of this feature article. For example this excludes the group of 2-dimensional crystals containing numerous oxides, including monolayers of $TiO_2$, $MoO_3$, $WO_3$, graphene oxide, silicates, perovskites; and hexagonal boron nitride, and elemental monolayers of Ge (germanene), Sn (stanine) and metals.

## 2. Graphene

Graphene can be visualised as an infinite single-atom-thick sheet of carbon arranged in a crystalline hexagonal lattice, and, in this regard, as a pseudo 'two-dimensional' material as it actually possesses a finite thickness. Intrinsically, a pure graphene lattice has zero magnetic moment and at 0 K behaves as an insulator. Above 0 K graphene is conducting and is magnetic due to the presence of spin-½ itinerant electrons. This weak form of magnetism is referred to as Pauli-paramagnetism. Graphene also shows a large diamagnetism similar to that of graphite[21, 59]. In reality, graphene can only exist if the sheets are of finite size[60], and hence must contain chemical or structural defects. Synthetic graphenes, for example those prepared by exfoliation[61], dispersion[62], chemical reduction[63], epitaxial growth[64], and chemical synthesis[65] exhibit a number of vacancies, lattice mismatching, adatoms, unpaired electrons, and mechanical strains (Figure 2a-d). These changes essentially modify not only the chemical properties of the material but also the electron spin-dependent physical properties.

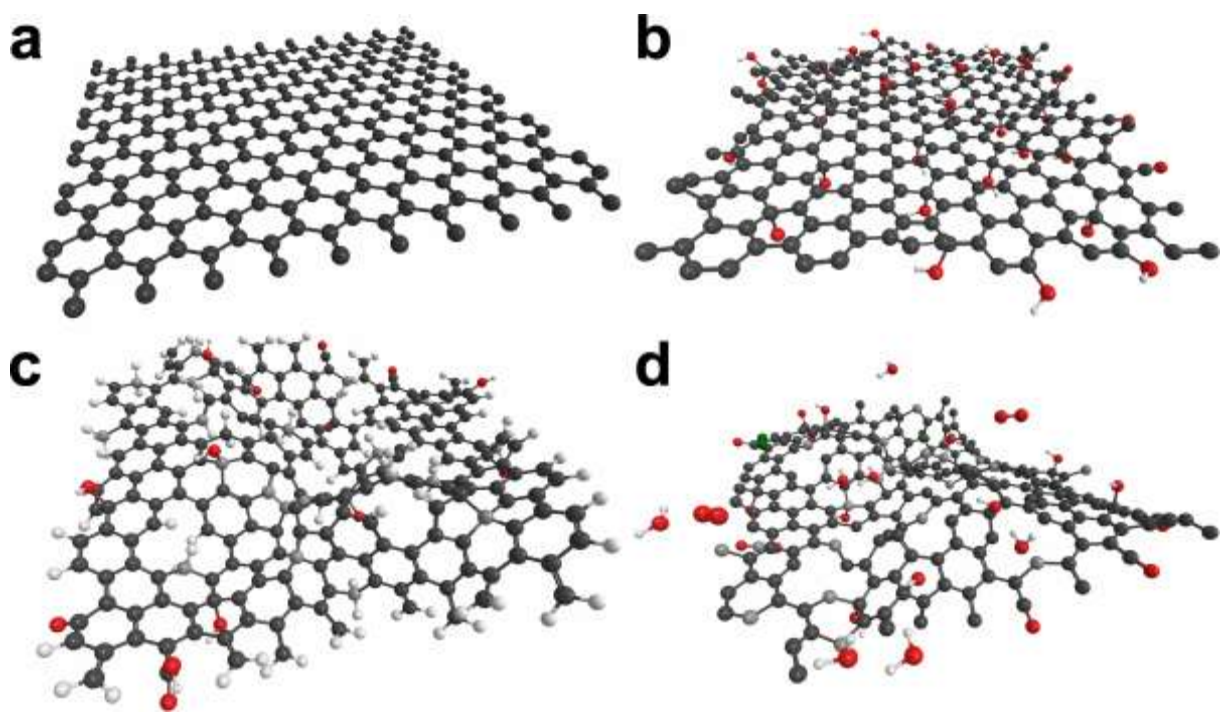

**Figure 2.** Molecular representations of various graphene materials. (a) Pristine graphene, (b) oxygenated graphene, (c) structurally modified, hydrogenated and oxygenated graphene, and

(d) structurally modified oxygenated graphene with adsorbates from air. Atom colours represent carbon (black/grey), oxygen (red), hydrogen (white), and nitrogen (green).

The presence of a magnetic moment close to or on the graphene surface disrupts the electronic structure of graphene[66] through a competition between valence and conduction electrons. Both unsaturated and saturated vacancies in graphene[67] as well as chemisorbed hydrogen[68], and other molecules including $HNO_3$ and $NO_2$[20], $O_2$[22], fluorine[17], and transitions metals[69], are known to bring a magnetic moment. Such a magnetic moment generally displays a spatial localisation of approximately 2.0 nm i.e. over several benzene rings. However, not all vacancies in graphene may be magnetic and theoretically it has been argued that there may exist a 'magic' number of defects that could result in tunable magnetism.[70]

There are a number of techniques used to detect spin magnetic moments of graphene, including SQUID magnetometry (superconducting quantum interference device), ESR, muon spin spectroscopy (μSR), magnetic force microscopy or scanning tunnelling microscopy, and spin transport measurements.

Room-temperature ferromagnetism in some bulk graphene samples detected using SQUID magnetometry has been reported to originate from defects[71]. However, in such samples prepared using oxidative precursors[72], magnetic impurities like $MnO_4^-$ in $KMNO_4$ could remain in the sample. Also, the signals obtained by SQUID magnetometry do not necessarily indicate the origin of the signal contribution and are at best semi-quantitative i.e. a very small amount of magnetic impurity may give a very large magnetic response, which is assumed to be distributed over the entire sample, as the magnetisation is calculated by dividing over the sample mass (hence, an accurate distribution of magnetic moments remains inconclusive). Multi-pronged efforts have been made to address this inconclusiveness by employing a number of techniques, including multi-frequency ESR[73], to probe for the spectroscopic response of magnetic impurities in graphene[74]. However, in the context of instrumental detection limits and in the absence of measurements performed locally, we stress that the absence of evidence for magnetic impurities is not necessarily an evidence of the absence of magnetic impurities. So, combined efforts in multi-disciplinary research in this field will fill such experimental gaps.

A commonly used technique for detecting and manipulating electron spin is electron spin resonance (ESR), whereby the electron dynamics in bulk quantities of magnetic graphene materials can be probed[8, 10]. ESR was used to determine that the conduction electron density in a synthetic graphene[65] could closely resembled that of 'pristine' graphene cleaved from natural graphite[23]. These measurements also provided a lower bound and record value of ~50 ns lifetime at 300 K for a clearly resolved itinerant electron spin in the synthetic graphene. This value was comparable to a conservative estimate of spin relaxation time resulting from the Elliot-Yafet mechanism of relaxation[75]. The record-long electron spin lifetime detected in this sample allowed for the spectroscopic monitoring of $O_2$ to within 1.1 nm of the graphene surface and demonstrated a clear correlation between the measured spin lifetime and $O_2$ proximity[22]. Recently, and rather counter-intuitively, it was shown by ESR that conducting metallic-like carbon nanospheres made up of short disordered graphitic fragments possessed an intrinsically long itinerant spin lifetime of ~175 ns at room temperature[4]. These times currently set the record for the intrinsic electron spin lifetime in conducting carbon nanomaterials at room temperature.

The magnetic moment of muons (3.2 times that of the proton) and the availability of 100% spin-polarised muon beams allowed the exploitation of muons as sensitive spin probes once implanted in graphene powders[76]. The 100% spin-polarised pulsed beam employed was particularly suited for the study of the muon spin evolution at long timescales (~15 μs) and for the detection of very low precession frequencies. Apart from the background of muons passing through the sample, two main contributions were observed in zero-field μSR data of the prepared graphene, independently of sample preparation: a Lorentzian relaxation of polarisation, experienced by a large fraction of the muons as a result of the isolated paramagnetic electrons located at unsaturated defects sites, and a damped oscillation, corresponding to the muon spin precession around a local magnetic field of the order of a few Gauss. In order to obtain the proper interpretation of this last signal, several muon decay functions were extensively tested and excluded, until the authors attributed the origin of the observed signal to the μ-H dipolar interaction[77]. And besides, the temperature dependence of the precession signal showed that it persisted up to 1250 K; a critical temperature that far exceeds the highest magnetic transition temperature ever reported in carbon materials. This effectively ruled out magnetic phases in the defective graphenes studied, including the presence of antiferromagnetic order, not easily detectable with conventional magnetometric techniques. However, as the authors state, the conclusion regarding the absence of magnetic ordering in synthetic graphenes did not concern other unambiguously proven cases.

In contrast to random defect distributions in bulk quantities of synthetic graphenes, nanofabrication techniques based on scanning tunnelling microscopy have been employed to prepare CVD-grown graphene nanoribbons with nanometre precision and well-defined crystallographic edge orientations[24]. This atomic-scale engineering of 'zigzag' edges (i.e. edge atoms from only one sublattice of the bipartite graphene lattice) gave rise to magnetic order. Edge states localised near the Fermi level render all zigzag graphene nanoribbons metallic[78]. These one-dimensional metallic edge states contain a high local density of states at the Fermi level and are magnetically unstable. However, by ordering the spins along the two ribbon edges with antiferromagnetic coupling between opposite edges, the energy of the system is lowered and a bandgap could be engineered (Figure 3). Consequently, the authors did not directly measure the magnetic signals: this would require bulk amounts of these nanoribbons. Rather, they were able to detect the signature of edge magnetism on individual graphene nanostructures by investigating their electronic structure. Therefore, upon increasing the ribbon width from <7 nm to >8 nm, a semiconductor-to-metal transition was observed, which indicated the switching of the magnetic coupling between opposite ribbon edges from the antiferromagnetic to the ferromagnetic configuration and was stable even at room temperature.

Local[79] and non-local[75, 80-83] spin transport devices have been used to measure magnetic moments in graphene by utilising magnetoresistance arising from spin accumulation via the spin Hall effect[84], which acts as the signal of spin transport (Figure 4). In the local geometry it is difficult to observe spin transport because of the presence of charge current between the spin injector and the spin detector, which produces a large spin-independent background signal. Non-local geometries have higher signal-to-noise ratio due to the absence of net charge flow between injector and detector. A key aspect of non-local resistance studies is the observation of Hanle spin precession,[83] which provides unambiguous proof of spin injection and transport in graphene. Although typical spin injection efficiencies in non-local geometries are low and range between 2-20%, efficiencies of greater than 60% have been obtained[85]. In general, both intrinsic mechanisms related to the spin-orbit coupling in band structure and helical scattering effects due to spin-orbit impurity potential scattering were postulated to contribute to the spin Hall effect.

Ferromagnetic metals would be ideal contacts for spin injection into semiconductors if it were not for the fundamentally negligible spin-polarisation between the two materials[86]. The presence of a tunnel barrier between the ferromagnetic metal and the semiconductor has been identified as a potential solution to this problem in the diffusive transport regime[87, 88]. Graphene has been predicted to behave as an efficient tunnelling barrier[89] and was shown to be an effective tunnel barrier in magnetic tunnel junction devices[87, 88, 90]. According to van't Erve et al.[88], graphene represented an effective tunnel barrier by meeting the following key material characteristics: uniform and planar in morphology with well-controlled thickness, minimal defect and trapped charge density, a low resistance-area product for minimal power consumption, and compatibility with both the ferromagnetic metal and the semiconductor that ensured minimal diffusion to and from the surrounding materials at the temperatures required for device processing. It was also meticulously shown by van't Erve et al.[88] that the large anisotropy of the electrical conductivity in graphene could be maintained if the contacts were designed so that the edges of the graphene were embedded in an insulator, preventing conduction through the graphene edge states that would of shorted out the tunnel barrier. Such magnetic tunnel junction configurations demonstrated that spin-polarised contacts integrating graphene could overcome the mismatch issue for electrical spin injection and detection in metal/semiconductor spintronics devices.

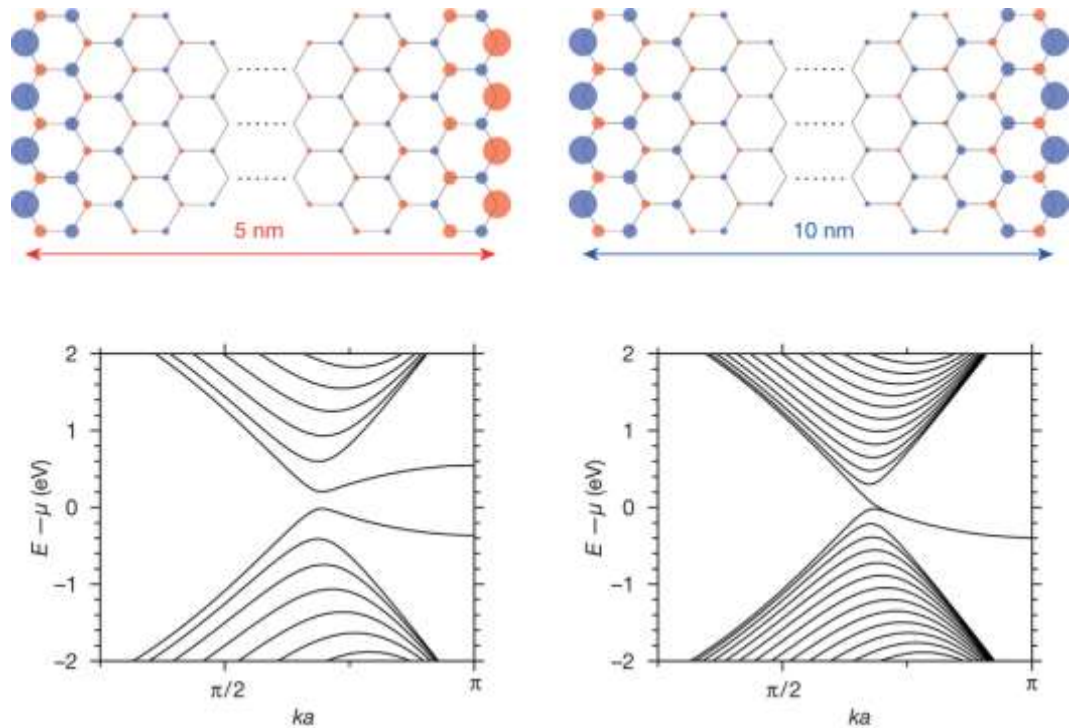

**Figure 3.** Correlating electronic and magnetic properties of zigzag graphene nanoribbons. Spin density distribution (↑, blue; ↓, red) in 5-nm-wide (left) and 10-nm-wide (right) zigzag graphene nanoribbons calculated in the mean field Hubbard model for $T$=300 K and $\Delta E_F \approx 100$ meV. The lower panels display the corresponding band structure, indicating that narrow zigzag ribbons are antiferromagnetic semiconductors, whereas the wider (>8 nm) zigzag ribbons display a ferromagnetic inter-edge coupling and no bandgap. $ka$ is wavenumber times the lattice constant. Reprinted by permission from Macmillan Publishers Ltd: Nature (reference [24]), copyright 2014.

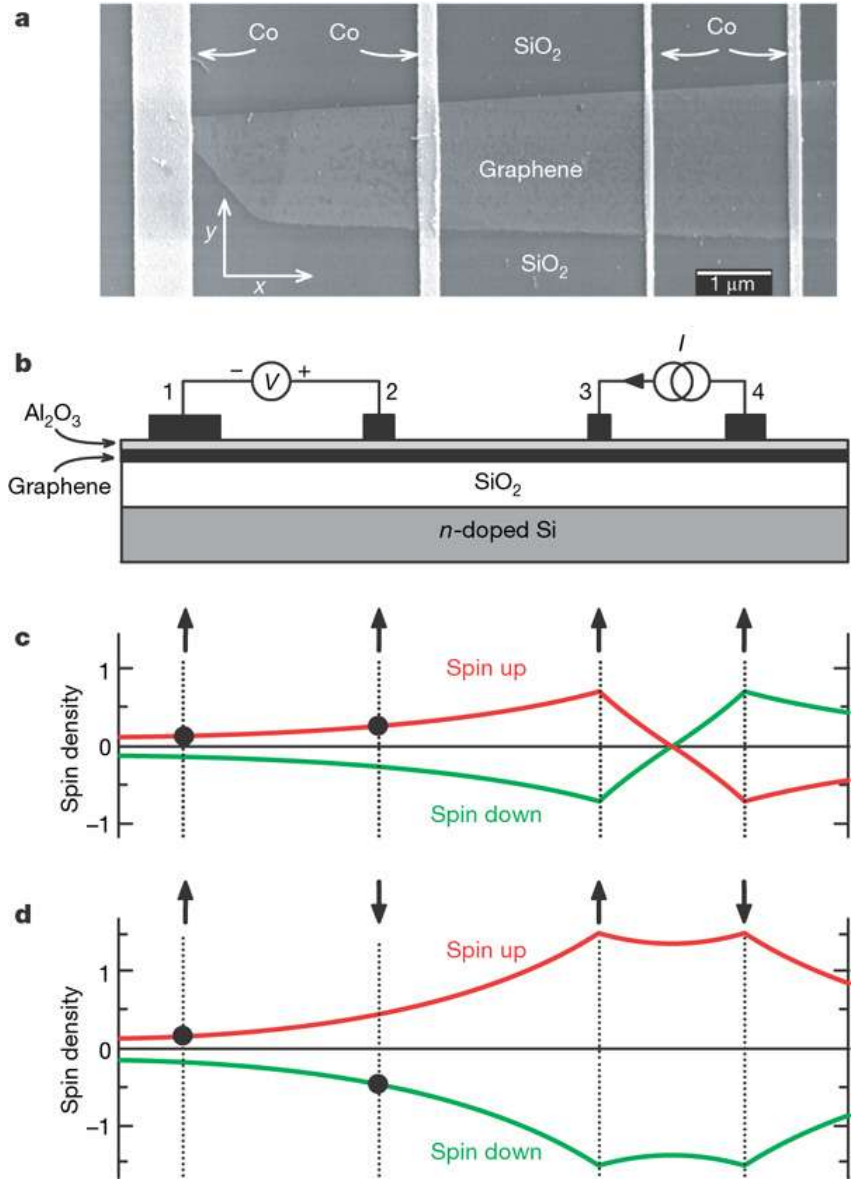

**Figure 4.** Spin transport in a four-terminal spin valve device. (a) Scanning electron micrograph of a four-terminal single-layer graphene spin valve. Cobalt electrodes (Co) are evaporated across a single-layer graphene strip prepared on a $SiO_2$ surface. (b) The non-local spin valve geometry. A current I is injected from electrode 3 through the $Al_2O_3$ barrier into graphene and is extracted at contact 4. The voltage difference is measured between contacts 2 and 1. The non-local resistance is $R_{non-local} = (V_+ - V_-)/I$. (c) Illustration of spin injection and spin diffusion for electrodes having parallel magnetisations. Injection of up spins by contact 3 results in an accumulation of spin-up electrons underneath contact 3, with a corresponding deficit of spin-down electrons. Owing to spin relaxation the spin density decays on a scale given by the spin relaxation length. The dots show the electric voltage measured by contacts 1 and 2 in the ideal case of 100% spin selectivity. A positive non-local resistance is measured. A larger positive signal can be obtained by reversing the magnetisation direction of contact 1. (d) Spin injection and spin diffusion for antiparallel magnetisations. The voltage contacts probe opposite spin directions, resulting in a negative non-local resistance. Reprinted by permission from Macmillan Publishers Ltd: Nature (reference [83]), copyright 2007.

Spin-orbit coupling is an essential interaction which causes spin decoherence[91]. The intrinsic spin relaxation of magnetic moments in graphene is dominated by spin-orbit coupling and is typically discussed in relation to the Elliott-Yafet[14] and D'yakonov-Perel'[13] spin-relaxation mechanisms depending on whether inversion symmetry is retained or broken, respectively. More recently, there has also been an attempt to unify the two theories on general grounds for a generic two-band system containing intra- and inter-band spin-orbit couplings[15]. Elliott showed using first-order time-dependent perturbation theory that there is a probability for an electron to flip its spin at a momentum scattering event. D'yakonov and Perel' showed that a lack of inversion symmetry in semiconductors results in an efficient relaxation mechanism that accounts for changes in spin direction between collisions with impurities. The D'yakonov-Perel' relaxation rate is proportional to the carrier mobility, unlike the Elliott-Yafet mechanism. The Elliot-Yafet rate is always smaller than the D'yakonov-Perel' rate contribution in good conductors like graphene.[15]

A conservative theoretical estimate of the Elliot-Yafet spin-flip time is ~50 ns (at resonant energies the spin-flip time equals the spin relaxation time)[92]. Mid-range theoretical estimates put the spin-relaxation time exceeding microseconds. D'yakonov-Perel' relaxation can occur in graphene when Bloch states are spin-split and the Rashba effect appears[93, 94] (which can be due to gating electric fields, ripples, space inversion symmetry broken by the substrate, or adatoms). However, these interactions may not necessarily be strong enough to be solely responsible for the spin lifetimes observed in graphene[92]: values obtained for graphene by spin-valve device measurements are of the order 0.5-2 ns at 300 K and 1-6 ns at 4 K[75, 80, 81, 95]. Hence, the discrepancy between experimentally and theoretically obtained values of spin lifetime is yet to be fully reconciled[56]. These observations indicate that although spin relaxation could be described by a number of intrinsic mechanisms, extrinsic interaction mechanisms must play an essential role.

Kochan et al.[96] proposed theoretically that short (100 ps) spin relaxation times in graphene could be partially due to resonant scattering of electrons by local magnetic moments. This resonant scattering mechanism is in qualitative agreement with experimental ESR observations of the detrimental effect of surface bound oxygen on the graphene electron spin lifetime.[22] Sosenko et al.[97] showed that spin valve and Hanle spin precession experiments suffered from additional spin loss due to the resistance mismatch between ferromagnetic electrodes and graphene, and not for instance, Coulomb scattering[98]. In a general sense, the intrinsic thinness of a material can lead to the breakdown of the common Elliot-Yafet and D'yakonov-Perel' models for spin relaxation. Identifying the spin-relaxation mechanisms that take place in graphene and at the material interface has allowed for a more strategic approach of increasing the spin lifetime towards the theoretical limits[80].

It must be noted that although the carbon isotope $^{13}C$, in contrast to $^{12}C$, possesses a nuclear magnetic moment that could induce electron spin dephasing in graphene, this effect is usually neglected due to the low abundance of $^{13}C$ in natural carbon allotropes (~1%). Indeed, a negligible effect of the hyperfine interaction in isotopically engineered $^{13}C$-graphene has been observed[99].

The degree to which localised spins interact with graphene substrates altering spin relaxation has recently been proposed[100]. In this case, the classical and quantum dynamics of molecular magnets on graphene were studied by measuring the response to an oscillating magnetic field. It was shown that while the static spin response remained unaltered, spin relaxation was heavily dominated by quantum tunnelling relaxation channels. Dipolar and hyperfine interactions introduced a local dynamic magnetic field distribution that altered the quantum tunnelling rate. Furthermore, intermolecular dipolar interactions were strongly reduced due to the large separation between the molecules on the graphene surface and magnetic shielding by graphene[26]. However, hyperfine interactions arising from the nuclei of the molecules on graphene were unaffected by the substrate. The only significant contribution from graphene vibrations was from the modulation of the anisotropy energy induced by long-wavelength acoustic phonons.

The topic of magnetic graphene remains open for discussion and controversial because graphene contains only $sp^2$ electrons. The very small magnetic signals detected experimentally, and reports of Curie temperatures far exceeding room temperature (i.e. we could expect strong ferromagnetism in carbon at room temperature) in modified graphenes[101] and experimental reports for ferromagnetism in other carbon allotropes including disordered hydrogenated carbon nanotubes[102] and quenched diamond-like carbon[103] need to be confirmed. A classic example of the controversy that has plagued the concept of magnetic carbon was the saga of room-temperature ferromagnetism in $C_{60}$[104] which spanned over 10 years and involved retractions by several authors, and the loss of a large number of resources. In a sense, this demonstrated the imperative need for a definitive account of the origin of magnetism in carbon materials, and on the other hand, it also highlighted that the impact of such results would represent a substantial paradigm shift in the field of magnetism in solids. The vigorous discussion and advocacy around magnetic carbons – in particular, graphene – still remains, and was spearheaded by the European Ferrocarbon Project involving a consortium of 7 European institutes[105].

At present, inquiries into the mutual dependence of defects, adatoms, and itinerant electrons and the degree to which they affect the graphene materials' magnetic properties is central to this open discussion. Hence, concerted efforts to understand the specific origins of magnetic behaviour in graphene materials remains of fundamental importance. Reports of ferromagnetism, antiferromagnetism, diamagnetism, and even suggestions of possible room-temperature superconductivity in graphene continue[106]. However, in the pristine state, graphene exhibits no signs of conventional spin polarisation and so far no experimental signature shows a ferromagnetic phase of graphene[107].

In order for graphene to be seriously considered as a potential candidate for spin-based devices, practical challenges need to be met. Even though the spin-orbit coupling in graphene may result in a relatively prolonged spin lifetime compared with other materials like silicon[56], and lead to many interesting phenomena including spin Hall effects[84, 108, 109], spin-dependent Klein tunnelling[94], and weak antilocalisation (which results in lower net resistivity)[110], it invariably introduces the need for band gap engineering to maintain magnetic stability. This will most likely need to be considered together with the deterministic incorporation of localised magnetic moments on graphene, which is a non-trivial task.

Nevertheless, the progress is encouraging. The itinerant spin lifetimes in graphene (1-50 ns) are approaching practical values of 100 ns at room temperature. The reported spin carrier mobilities at room temperature are high and the spin diffusion lengths are typically of the order of 100 nm. These fundamental properties are compatible with device dimensions below 100 nm. Understanding of the component interactions and influences in graphene-based spintronics has led to more robust and efficient device design. Graphene may now be confidently reproduced from non-graphitic precursors with well-defined dimensions. Research has now moved beyond graphene synthesis to address the need of developing efficient methods to coherently controlling the quantum characteristics of components in scalable graphene-based devices.

3. Phosphorene

Black phosphorus is an elemental layered material with an inter-layer spacing of ~0.53 nm and an orthorhombic crystal structure with a lattice constant along the *z*-direction of 1.05 nm[111]. Individual layers in black phosphorus are referred to as phosphorene. Each phosphorus atom in phosphorene is connected to three adjacent phosphorus atoms to form a linked ring structure, with each ring consisting of six phosphorus atoms. The 'puckered', i.e tightly corrugated, structure results in optical anisotropy[112] and a reduced in-plane symmetry with respect to the hexagonal array found in graphene (Figure 5).

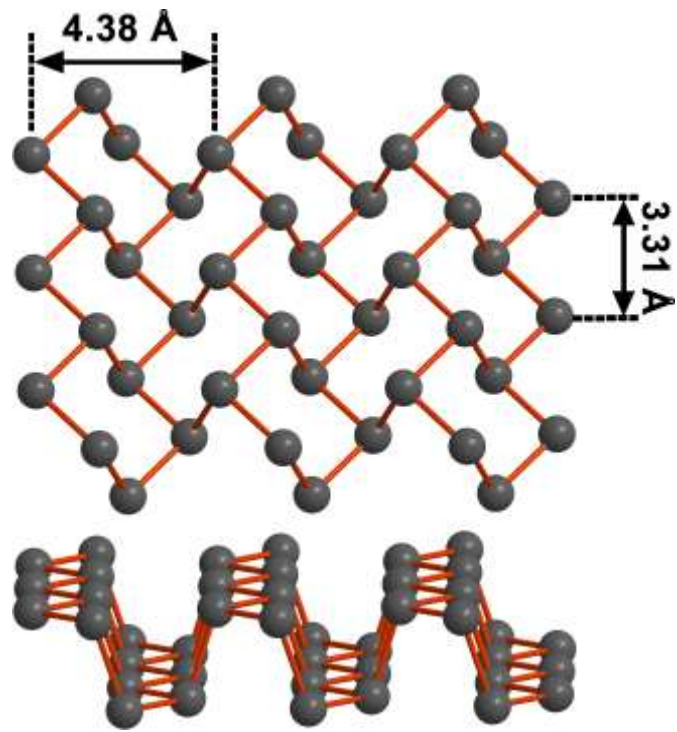

**Figure 5.** Structure of monolayer black phosphorus (phosphorene). Lattice parameters $a$ = 4.38 Å and $b$ = 3.31 Å are shown in the view looking down onto the sheet. The lower structure is a side view showing the puckered geometry.

Black phosphorus is a p-type semiconductor with a direct band gap of 0.33 eV which has been predicted to increase up to ~2 eV in phosphorene[113, 114]. Very recently, it was found that potassium doping the surface of black phosphorus closes its band gap, producing a Dirac semimetal state with linear dispersion in the armchair direction and a quadratic one in the zigzag direction[42, 115]. The possibility of a tunable direct band gap in black phosphorus layered materials has given rise to potential applications in photovoltaics,[116-118] photocatalysis[119], transistors[118, 120, 121, 122], and gas sensors[123].

Few-layer black phosphorus field-effect devices have been fabricated on Si/SiO$_2$[122]. This work highlighted a major challenge for the nominal[124] fabrication and operation of devices based on layered black phosphorus: after only 20 minutes, the surface roughness measured by atomic force microscopy of the black phosphorus layers more than doubled from 427 pm to 977 pm, demonstrating rapid oxidation and poor material stability. Other works by Kang et al.[125] also observed similar material degradation (Figure 6). However, since then, there have been promising advances to address this issue: the first example of an air-stable layered phosphorene FET was demonstrated that comprised of high-quality hexagonal boron nitride encapsulating layer and monolayer graphene electrodes[120]. This simple, all 2-dimensional layered material architecture resulted in hysteresis-free transport measurements with the added benefit that it could be generally applicable for other sensitive two-dimensional crystals. Conductivity could also be retained upon device cooling to a greater extent than air-exposed samples and also when compared to other Ti/Au contacted devices[126], however, the poor chemical stability resulted in a trade-off with up to an order-of-magnitude decrease in carrier mobility and subsequent device designs have attempted to address this issue[29, 43, 44, 127, 128].

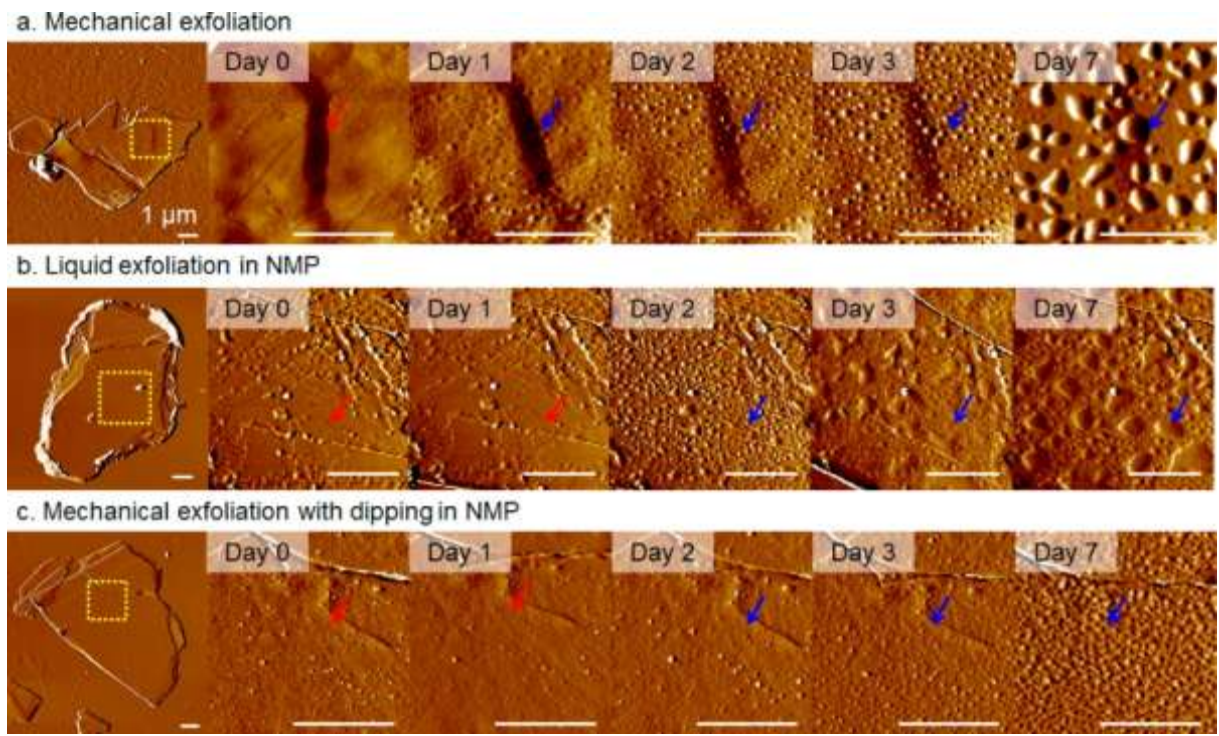

**Figure 6.** Black phosphorus monolayer stability. FM amplitude images (amplitude scale: -5 to 5 nm (top left), -1 to 1 nm (magnified images)) of black phosphorus flakes prepared by (a) mechanical exfoliation, (b) solvent exfoliation in *N*-methylpyrrolidone (NMP), and (c) mechanical exfoliation followed by 1 h submersion in NMP. The leftmost image shows the entire flake, and the images progressing to the right show magnified views immediately after exfoliation up to 7 days in ambient conditions. Structural deformations (*i.e.*, apparent bubbles) are observable on the surface of the mechanically exfoliated sample after 1 day and on the rest of the samples after 2 days. Red and blue arrows indicate the same position on the flakes before and after the appearance of bubbles, respectively. All flakes are thicker than 150 nm, and all scale bars are 1 μm. Reprinted with permission from Ref.[125] Copyright 2015 American Chemical Society.

Spintronic devices require spin currents to be transported over practical distances, and the generation and detection of tuneable spin currents, which could be done using magnetic semiconductors with a high carrier mobility[7]. The velocity of charge carriers in layered black phosphorus[42] has been found to be about half of that in graphene[66]. Typical room temperature charge mobilities in layered black phosphorus materials have been reported from 40 $cm^2 V^{-1} s^{-1}$ to ~1350 $cm^2 V^{-1} s^{-1}$ in various devices[29, 43, 44, 117, 118, 120, 121, 122, 125-129], and as high as 6000 $cm^2 V^{-1} s^{-1}$ below 30 K,[44] which allowed for the observation of the quantum Hall effect (Figure 7). However, these values are still much lower than the calculated room-temperature value of 10000 $cm^2 V^{-1} s^{-1}$ for few-layered black phosphorus[114] and the 65000 $cm^2 V^{-1} s^{-1}$ obtained experimentally on bulk single crystals[130], possibly due to the effects of electron-phonon scattering, material stability, charge traps, impurities, substrates, and high Schottky barrier heights[29, 43, 120, 127, 131, 132]. And even though the mobility values reported are still much lower than that found in graphene (15000 $cm^2 V^{-1} s^{-1}$)[66], they compare well with the best values found in single-layer $MoS_2$ transistors (~200 to 500 $cm^2\ V^{-1}\ s^{-1}$)[133] and graphene nanoribbons (~100 to 200 $cm^2\ V^{-1}\ s^{-1}$)[134].

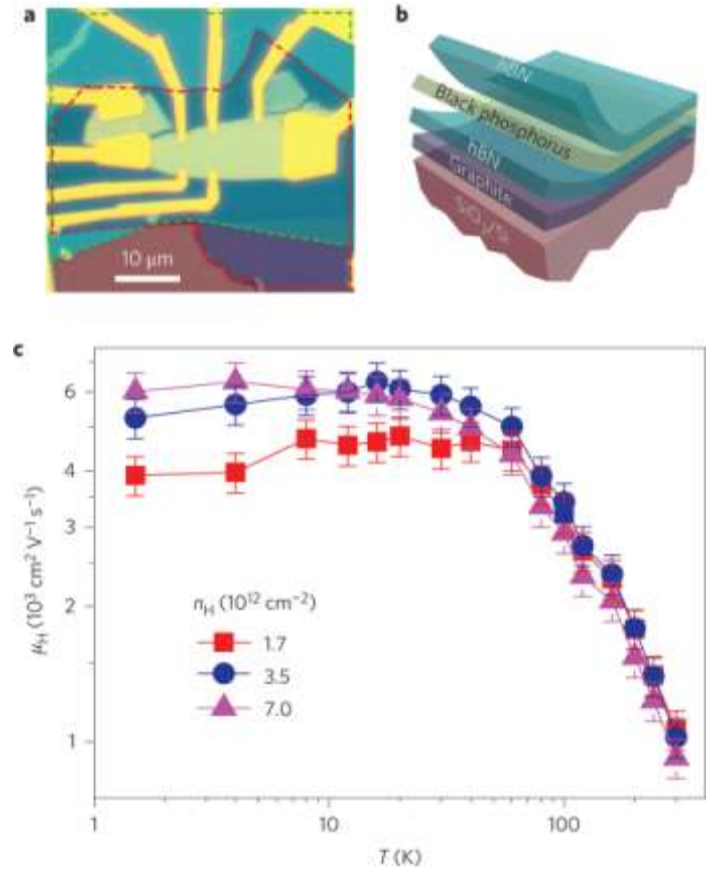

**Figure 7.** The device structure and mobility characterisation of a black phosphorus 2-dimensional hall gas. (a) Optical image of a black phosphorus/hBN/graphite heterostructure with graphite serving as the back gate. The boundaries of the hBN and graphite areas are marked by green and red broken lines, respectively. A layer of hBN (not shown) is later deposited on top to protect the black phosphorus from degradation in air. (b) Schematic three-dimensional view of the complete heterostructure stack of the device in (a). (a) Hall mobility $\mu_H$ as a function of temperature measured at varying hole carrier densities. The vertical error bars represent uncertainties in determining the sheet conductance from the measured sample resistance as a result of the irregular sample geometry. Data were obtained from the device shown in (a). Reprinted by permission from Macmillan Publishers Ltd: Nature Nanotechnology (reference [44]), copyright 2016.

A number of works published independently and almost simultaneously[29, 43, 127, 128] showed that a 2-dimensional electron gas could be induced on a black phosphorus surface using a gate electric field while probing for Shubnikov-de Haas oscillations in the magneto-resistance, and that carriers were mostly confined within approximately two atomic layers[29]. A weak localisation effect was observed in disordered black phosphorus[43]. It was experimentally noted[29, 43, 127, 128] that the non-trivial Berry's phase[135] does not exist in pristine black phosphorus[136]. Results of magnetotransport studies were consistent with holes in black phosphorus being Schrödinger fermions devoid of pseudospin.

The theoretical value of effective mass for the in-plane cyclotron motion of holes in few-layer black phosphorus was found to be ~0.35 $m_0$ (where $m_0$ is the mass of a bare electron)[114]. Experimentally obtained values[29, 43, 127, 128] were found to range between 0.27 $m_0$ and 0.36 $m_0$, with the cyclotron mass increasing as the sample thickness approached a monolayer, which consequently results in a decreased carrier mobility. The cyclotron mass of electrons[29] was found to be 0.47 $m_0$.

The carrier lifetime for electrons and holes in 2-dimensional black phosphorus were found to be on the order of ~120 ps[29, 127]. The promising value of carrier lifetime represents a lower bound, as it neglects the effects of spin-orbit coupling[137]. Following from the introduction of Landau levels by Zeeman splitting[136], simple experimental estimates of the temperature and magnetic field-independent g-factor for holes and electrons gave an upper bound $g<2.94$ and a lower bound $g>2.13$, respectively[29, 128]. The appreciable deviation from $g\sim2.00$ indicated that contributions of spin-orbit coupling and lattice anisotropy could already be considerable[114, 137].

Although electron and hole doping of layered black phosphorus materials in early experiments showed promising signs for its use in spintronics, semiconducting layered black phosphorus materials are intrinsically nonmagnetic[18, 42, 138-143]. A number of examples systematically exploring chemically[18, 19, 132, 139, 141, 144-146] and structurally[138, 140, 142, 147, 148, 149] induced magnetic states in layered black phosphorus and phosphorene have been proposed by *ab initio* calculations.

The 3$d$ transition metals (Ti, V, Cr, Mn, Fe, Co, Ni) are expected to produce a diverse variety of magnetic properties in phosphorene[19, 141, 144-146]. The predicted magnetic moments in phosphorene calculated by introducing transition metals are mainly attributed to the $d$ orbitals of the investigated transition metals. When adsorbed on phosphorene, the magnetic moments of transition metal adatoms are reduced. It has been found that transition metal dopants such as Mn and Fe could bind with P vacancies in phosphorene to form a defect complex that may be ferromagnetic[141]. Hashmi et al.[145] suggested that the most promising transition metal dopants in the context of creating dilute magnetic semiconductors were Ti, Cr, and Mn, because the spin polarised state was achieved with a finite-size band gap and with minimal suppression of the magnitude of bulk magnetic moment.

The first-principles study of metals on phosphorene by Kulish et al.[19] showed that metals, such as, Ni, Cu, Ag, Pd and Pt formed electrically inactive states which were located entirely outside the band gap, transition metals including Ti, V, Cr, Mn, Fe and Co formed spin-polarised states across the entire band gap region as well as peaks in the valence and conduction bands, and the alkali metals like Li, Na and K preserved the original electronic structure of phosphorene but shifted the Fermi level to the conduction band, unlike Au and Pt adatoms that acted as electron acceptors. Hu et al.[144] also predicted similar behaviour in ~2% metal doped phosphorene, however, a spin-polarised band gap structure could be obtained in Fe, Co, and Au doped systems by tuning the metals' aggregating behaviour. While Sui et al.[146] found that a small biaxial strain could induce a magnetic transition from a low-spin to a high-spin state in phosphorene decorated by Sc, V, or Mn.

Theoretical methods have also been used to look into the magnetic properties of non-magnetic[139] and non-metal[18] doped phosphorene. In particular, the work of Zheng et al.[18] showed that the substitutional doping of H, F, Cl, Br, I, B, N, As, $C^-$, $Si^-$, $S^+$, or $Se^+$ could not induce magnetism in a phosphorene monolayer due to the saturation or pairing of valence electrons of the dopants and their neighbouring P atoms, whereas the ground states of neutral C, Si, O, S, or Se doped systems were magnetic due to the appearance of an unpaired valence electron of C and Si, or the formation of a nonbonding 3p electron of a neighbouring P atom around O, S and Se. Furthermore, the magnetic coupling between the moments induced by two Si, O, S, or Se were antiferromagnetic and the coupling was attributed to the hybridisation interaction involving polarised electrons, whereas the coupling between the moments induced by two C atoms was weak. And although doping introduced structural deformation, a promising aspect of the work showed that the systems were actually energetically stable.

Simulated point defects – more generally structural modifications – in phosphorene, can give rise to magnetic states[138, 142, 147, 149]. Specifically, single-vacancy and double-vacancy defects can introduce unoccupied localised states into the band gap of phosphorene. Specifically, the 5-9 single vacancy (i.e. a single vacancy with adjacent nonagon and pentagon ring) was found to give a ferromagnetic state due to the presence of a dangling bond in a specific point defect made up of a $P_{60}$ cluster[147]. Zhu et al.[149] showed that both ferromagnetic and antiferromagnetic states were possible in zigzag-edge phosphorene nanoribbons and that the magnetism arose from dangling bonds as well as edge-localised π-orbitals, while the oxygen-saturated ribbons gave rise to magnetic ground states due to the weak P-O bond in the ribbon plane between the $p_z$-orbitals of the edge O and P atoms. Hashmi et al.[138] also explored the possibility of long-range magnetic ordering originating from edges in porous phosphorene. The self-passivated pore geometry showed a nonmagnetic state while the pore geometry with dangling bonds at two zigzag edges preferred an antiferromagnetic state that changed to long-range ferromagnetic ordering with the introduction of an external electric field (as the energy difference between the two states was suppressed). The magnetic tails along the armchair direction were found to be delocalised and formed long-range antiferromagnetic ordering which was preserved by edge passivation with oxygen. From these works, it is now known that electronic states at zigzag edges in phosphorene could result in spin polarisation while retaining a band gap.

Black phosphorus is an exciting rediscovery for the materials science community and it has quickly become the subject of significant theoretical and experimental investigations. Although, a trade-off has quickly emerged when considering the doping and structural modification of phosphorene between obtaining desirable magnetic states and trying to retain a finite band gap, structural stability, and the extent of reactivity (i.e. associated binding energies and activation energies to diffusion). And despite the basic universal approach taken so far to fabricate phosphorene-based devices and to simulate possible electronic and magnetic states, the developments towards phosphorene technologies have come with a set of unique challenges, the least of which have concerned the hazardous nature of chemically processing phosphorus materials[125, 150].

The measured values of carrier lifetimes in phosphorene (<<1 ns) are currently well below practical values of 100 ns at room temperature. The reported carrier mobilities are low and the spin diffusion lengths are expected to be on the order of a few nanometres. These fundamental properties are currently impractical for spin-transport devices. However, the learnings from graphene device fabrication has allowed for a rapid improvement in phosphorene-based device design which is expected to result in more robust components. The field of phosphorene research continues to grow rapidly in large part due to the ease of layered material sample preparation and the ability to translate and integrate the layered material samples into conventional characterisation tools and device fabrication methods. Together with the intrinsic properties of a tuneable finite band gap, the potential for high carrier mobilities, and the possibility of dilute magnetism, the field of phosphorene research offers a versatile option for developing spintronics devices.

## 4. Silicene

Silicene is a monolayer of undulating, hexagonally arranged silicon atoms. The vertical displacement of two silicon atoms in opposite directions in the unit cell provides a more stable configuration than the planar one.[151, 152] The magnitude of the layer buckling has been calculated to result in different electronic band structures around the Fermi energy level: the low (equilibrium) buckling of 0.44 Å could result in semimetallic sheets, while a high buckling of 2.13 Å in metallic ones.[151] The buckled structure also arises from interactions between the silicene layer and the substrate.[153] For low-buckled silicene geometries, a lattice constant of $a = 3.86$ Å and nearest neighbour Si-Si distance of $d = 2.28$ Å are typical values (Figure 8).[39]

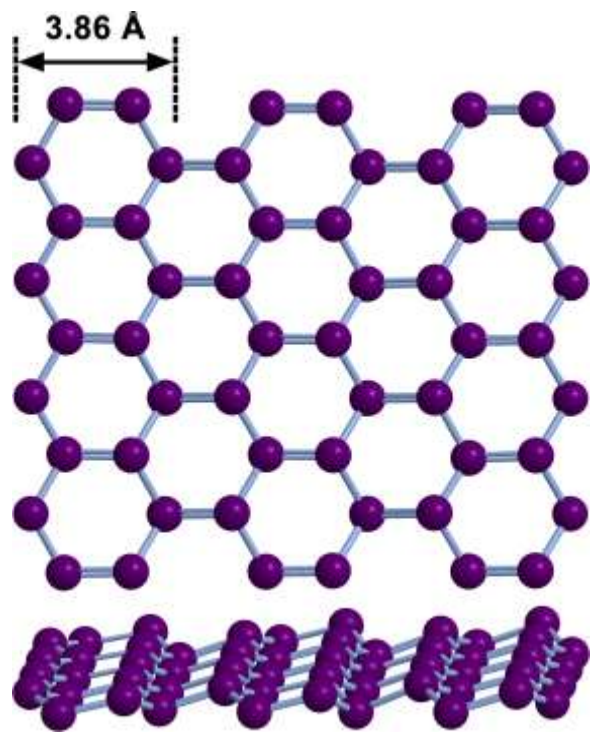

**Figure 8.** Structure of silicene. Lattice parameter *a* = 3.86 Å is shown in the view looking down onto the sheet. The lower structure is a side view showing the buckled geometry.

A naturally occurring layered parent silicon crystal, like that for which graphite is cleaved to obtain graphene and black phosphorus for phosphorene, has not yet been discovered. Rather, silicene can be produced by the epitaxial growth of silicon on various substrates including Ag(111) (ref. [154, 155-158]), Ag(100) (ref. [155]), Ag(110) (ref. [155, 159]), ZrB$_2$ (ref. [160]), Ir (ref. [153]). On Ag(100) and Ag(110), silicene grows as nanoribbons, while on Ag(111), silicene grows as sheets (Figure 9). The almost perfect lattice matching with the silicon honeycomb lattice and Ag(111) and a low tendency to form an Ag-Si alloy, makes Ag(111) an ideal substrate for the growth of silicene. However, silicene can be difficult to reproduce.[156, 161] The silicene sheet can present different orientations relative to the Ag(111) surface on varying the substrate temperature, giving rise to different superstructures with respect to the substrate, including 4×4 (ref. [155]), (2√3×2√3)R30° (ref. [162]), and (√13×√13)R13.9° (ref. [161]). It should be noted when attempting to structurally characterise silicene crystals, that corrugated silicon honeycomb lattices can also occur for silicides like CaSi$_2$, as they are also precursors to chemically exfoliated silicene flakes.[163]

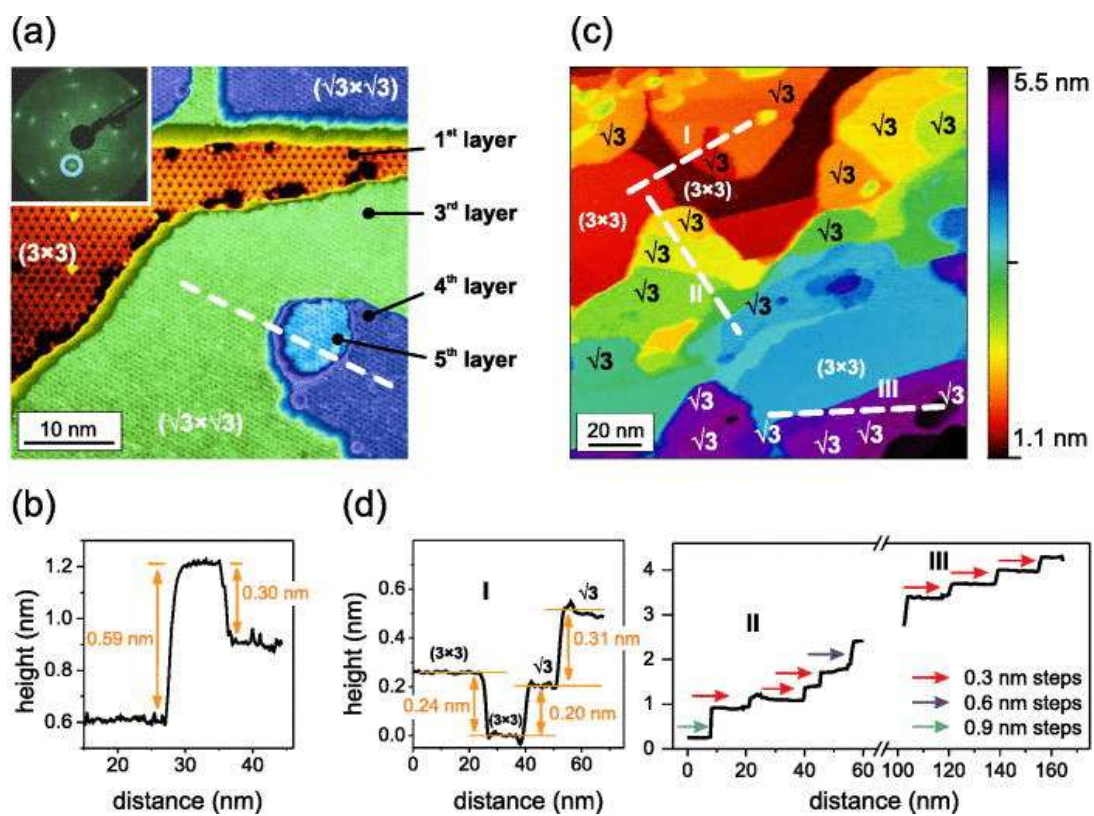

**Figure 9.** Scanning tunnelling microscope (STM) topography of silicene multi-layered sheets on Ag(111). 3 well-ordered terraces can be seen showing the (√3×√3)R30° arrangement. The 1st (3×3) silicene layer visible at the bottom. Inset: LEED pattern of the surface where the (1/3,1/3) type superstructure spots of the (√3×√3)R30° silicene are circled in blue. (b) Line profile along the dashed white line in (a). (c) STM topography, after greater deposition of Si onto the (3×3) silicene layer. (d) Line profile along the dashed white lines (indicated as I, II,

and III) in (c). Reprinted from Ref. [157], with the permission of AIP Publishing, copyright 2014.

Acun et al.[164] used low energy electron microscopy to record real-time images of changes in surface topography during the growth of silicene. They found that the presence of silicon adatoms on top of the silicene easily perturbed the $sp^2$ bonding hybridisation configuration causing a transition to the $sp^3$ bonding hybridised state. This is distinctly different from carbon, where the $sp^3$ bonding hybridised state (i.e. diamond) is less stable than $sp^2$ bonding hybridised carbon in graphene and graphite. They also found that the temperature-induced phase transition for silicene on Ag(111) under high vacuums occurred at approximately 600 K; that of graphene occurs at markedly higher temperatures.[165] Therefore, epitaxially grown silicene would not be expected to survive isolation from its parent substrate while exposed to air below ~100 °C (ref. [35]). Indeed, Molle et al.[158] found that a dramatic change in the silicene composition can be deduced already after 3 min exposure to air. This was overcome by encapsulating the silicene layer 4×4 and √3×√3 superstructures with a 7-nm-thick grown Al film that was sacrificially oxidised forming an $Al_2O_3$/Al/silicene/Ag(111) heterostructure. Although, it has been claimed by De Podova et al.[166] that thick epitaxial multilayer silicene films with a (√3×√3)R30° surface structure show only mild surface oxidation after 24 h in air without any protective $Al_2O_3$ capping. Regardless of the extent of oxidation, it is clear that epitaxially produced silicene is unstable in air.

However, epitaxially grown silicene has been found to undergo only a scarce degree of oxidation to $SiO_2$ upon exposure to $O_2$. This oxidation is limited to structural defects or domain boundaries,[158] with the oxidation process starting at very high $O_2$ exposures – about 100 times higher than on a clean Si(111) surface[167]. Interestingly, silicene monolayers grown on Ag(111) surfaces have been shown to demonstrate a band gap that is tunable from semimetallic to semiconducting type by oxygen adatoms. With the use of low-temperature scanning tunnelling microscopy, Du et al.[41] found that the adsorption configurations and amounts of oxygen adatoms on the silicene surface could be exploited for band-gap engineering. A promising outcome of this work was that the silicene monolayers retained their structures even when fully covered by oxygen adatoms, demonstrating a feasible approach to tuning the band gap of silicene.

Tao et al.[168] built a silicene back-gate field-effect transistor device that operated at room temperature. They use what was dubbed the SEDNE process, short for 'silicene encapsulated delamination with native electrode'. This process involved using a thin film of Ag(111) on a mica substrate to grow the silicene, instead of using single-crystal Ag substrates. The grown silicene formed mixed overlayers of 4×4, √13×√13, and (2√3×2√3) superstructures (Figure 10). The monolayer silicene was encapsulated with $Al_2O_3$ to prevent degradation during delamination of the silver–silicene–alumina sandwich and the upside-down transfer onto a device substrate ($SiO_2$ on highly doped silicon). The team patterned the silicene channel and used the native Ag(111) film for the source/drain electrodes, using a specially devised etchant to avoid the rapid degradation/oxidation of silicene.

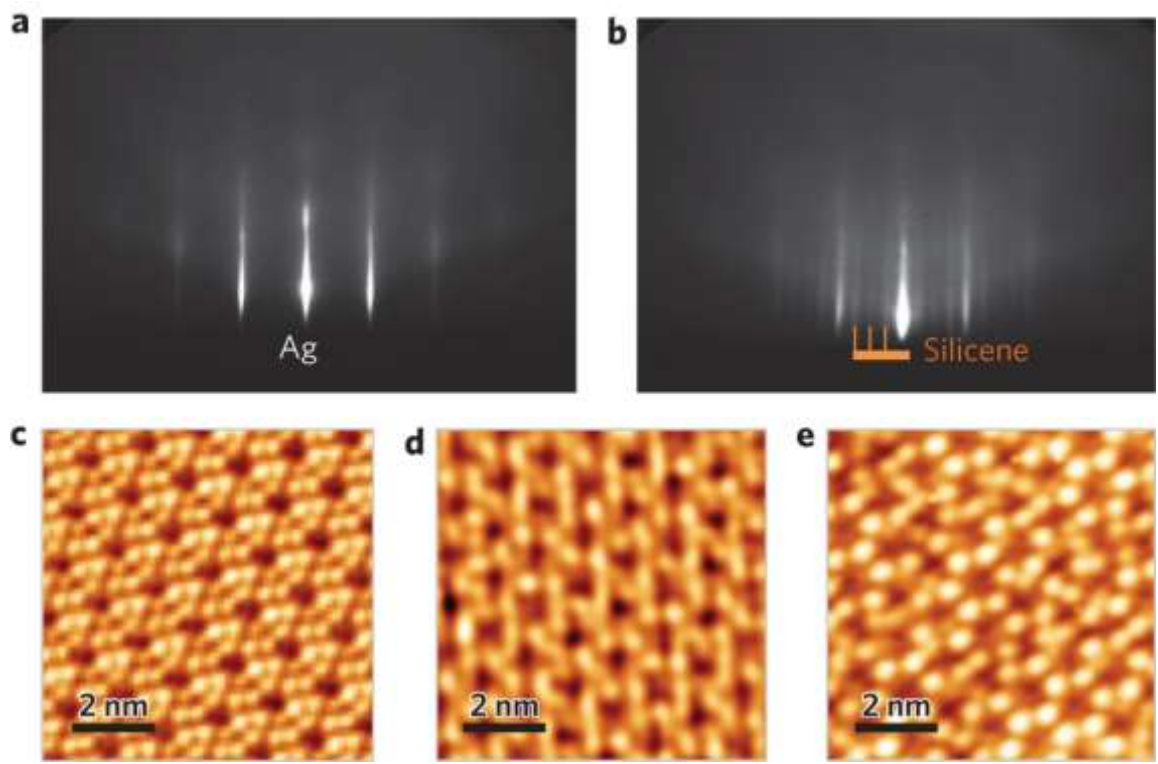

**Figure 10.** Grown silicene forming mixed overlayer superstructures. (a) Real-time reflection high-energy electron diffraction (RHEED) on Ag(111) and (b) silicene on Ag(111). *In situ* STM showing three main Si overlayers: (c) (4 × 4), (d) ($\sqrt{13} \times \sqrt{13}$) and (e) ($2\sqrt{3} \times 2\sqrt{3}$) superstructures. The STM images are 10 × 10 nm$^2$. Reprinted by permission from Macmillan Publishers Ltd: Nature Nanotechnology (reference [168]), copyright 2015.

The silicene devices showed Dirac-like ambipolar charge transport similar to graphene FETs, a high current modulation by the gate voltage (on/off ratio ~10) compared with graphene, and the possible existence of a band gap of ~210 meV perhaps due to the interaction with substrates. The measured mobility for both electrons and holes was ~100 cm$^2$ V$^{-1}$ s$^{-1}$ for both the mixed overlayer phase and single $2\sqrt{3}\times2\sqrt{3}$ phase devices, which is quite low with respect to estimated intrinsic values for silicene (~2 × 10$^5$ cm$^2$ V$^{-1}$ s$^{-1}$)[30]. The low mobility was attributed to strong scattering from acoustic phonons due to the symmetry-breaking buckled nature of the silicene sheet and to grain boundary scattering. Although the mobility values were about as promising as those found in MoS$_2$ (ref. [133]) and phosphorene[120], the work showed that stable silicene-based devices could be fabricated. However, electrical measurements could be performed only for a few minutes under ambient conditions before the monolayer silicene channel degraded to an amorphous insulator. Higher mobility and better device performance, in the future, could possibly be achieved by tuning the band gap by surface adsorption of molecules[41] or by using multilayer silicene[166].

Future devices based on silicene could integrate the concepts of valleytronics and spintronics using both coupling between multiple extrema of the band structure and the internal degree of freedom of spin, respectively. Due to strong spin-orbit coupling, inversion symmetry-breaking, and the band structure response of silicene to an external perpendicular electric field, spin- and valley-polarised charge carriers have been calculated to appear due to the $n = 0$ Landau level splitting between four distinct spin and valley energies.[169, 170] The band structure of silicene is similar to that of graphene in that the conduction and valence edges occur at the corners (K and K' points) of the Brillouin zone.[171] Silicene is comprised of silicon – an atomically heavier element than the carbon in graphene – and presents a larger spin-orbit coupling than graphene.[109, 172] Spin-orbit coupling generates spin polarisation.[173] The entanglement between spin and orbital degrees of freedom due to spin-orbit coupling reduces the degree of spin polarisation of spin-split states in non-magnetic semiconductors. In silicene, it has been calculated that the strong spin-orbit coupling could open band gaps at the K- and K'-points leading to the possibility of detectable quantum spin Hall effect below 35 K.[39, 170, 172] These gaps can be tuned with external electric fields perpendicular to the plane, which breaks the inversion symmetry of the system due to the presence of undulations in the honeycomb structure.[38, 171] If such spin degrees of freedom in silicene were available, or an ability to conduct charge and spin in gapless edge states without dissipation at the sample boundaries, it would make silicene a promising material for spintronic applications.

Tsai et al.[40] proposed a high-efficiency and tunable silicene spin-filter and spin separator that takes advantage of bulk charge carriers rather than the edge current in quantum spin Hall systems. Their first-principles computations showed that the band structure of gated silicene with zigzag edges harbours two nearly 100% spin-polarised Dirac cones at the K-points. These bulk states with non-spin-flip scattering processes were the crucial ingredients for the high-efficiency spin polarisation. The advantages here would be that the silicene spin filter would be robust against weak disorder and edge imperfections and that their device may work at 97 K, which is above the boiling point of liquid nitrogen. Although such silicon-based devices, in principal, could be feasible and the use of silicon is attractive industrially, the Ge, Sn and Pb counterparts of silicene were shown to have similar properties, and that their larger spin-orbit coupling resulted in larger energy differences between the spin-split states making them better suited for room-temperature applications.

Xu et al.[174] performed first-principles quantum transport calculations and predicted a magnetoresistance in zigzag silicene nanoribbons of up to 1960% at 300 K through the switching of the edge spin directions. They found that the spin-filter efficiency of both the antiferromagnetic and ferromagnetic silicene nanoribbons was sign-changeable with the bias voltage and presented a feasible approach to a prototype spin-valve device. These calculations and the device concept in large part were analogous to earlier works demonstrating very large magnetoresistance in graphene nanoribbons[175]. This work agreed well with earlier calculations by Cahangirov et al.[151] in that silicene nanoribbons display antiferromagnetically-coupled edge state configurations that are slightly lower in energy than ferromagnetic coupling (metallic state). In the presence of a magnetic field, the silicene nanoribbons can then switch between magnetic configurations leading to the expectation of magnetoresistance due to the current different between semiconducting and metallic states. However, as Cahangirov et al.[151] note, such magnetic properties depended strongly on the size, geometry, and chemistry of the silicene nanoribbons and the energy differences between magnetic states need to be carefully considered within the accuracy limits of the *ab initio* calculation methods employed and the inclusion of effects like spin-orbit coupling.

The direct measurement of spin-carrier relaxation times and coherence length of silicene would be necessary to assess the feasibility of silicene-based spintronic devices. The reported mobilities for epitaxial grown silicene are low. Silicene is chemically unstable under ambient conditions which makes accurate measurements of intrinsic properties unreliable and difficult. There are limited methods to produce silicene. With emerging methods to silicene production and handling, solutions could emerge to overcome challenges related to stability and device integration. Silicene presents only as a 'superficially' ideal material for compatibility with silicon electronics as the production methods to silicene currently differ to the methods producing silicon wafers and the two-dimensional material properties would necessarily, in a non-trivial sense, need to differ from the bulk. Charge carriers in 3-dimensional silicon nanomaterials have relaxation times less than 300 ps and spin coherence lengths of 200 to 350 nm (ref. [176]). If the mean free paths and spin-coherence lengths of carriers in silicene are similar to those in bulk silicon, silicene channels would need to be less than ~10 nm in length to perform room-temperature spin transport measurements. Also, the energy difference between magnetic states in silicene geometries may require temperatures below ~100 K to avoid spontaneous transitions that would void the device operation[174]. Nevertheless, we could expect much longer spin relaxation times at lower temperatures (e.g. 4 K), which would provide greater flexibility in device fabrication and measurement.

## 5. Transition Metal Dichalcogenides (TMDs)

The discovery of graphene placed significant attention on shuttling between a materials' physical properties when thinning a bulk crystal of macroscopic dimensions down to a few atomic layers. Like graphite, transition metal dichalcogenide (TMD) bulk crystals consist of monolayers bound to each other by Van der Waals attraction. In the TMD monolayers of the type $MX_2$, where *M* is a transition metal atom (e.g. Mo, W, etc.) and *X* is a chalcogen atom (O, S, Se, Te, Po), one layer of *M* atoms is sandwiched between two layers of *X* atoms and the structure has no inversion centre (Figure 11). Such TMD monolayers have a direct band gap, and hold potential for use in electronics as transistors and in optics as emitters and detectors.[177, 178] The emerging new degree of freedom of charge carriers due to the lost inversion symmetry is the *k*-valley index, which could be harnessed in electronic devices[179-181, 182] similarly to the case of graphene[183]. Moreover due to the strong spin orbit coupling robust type-II Weyl semimetal topological phase is predicted to exist which promotes TMDs as good candidates for electronics and spintronic applications.[184]

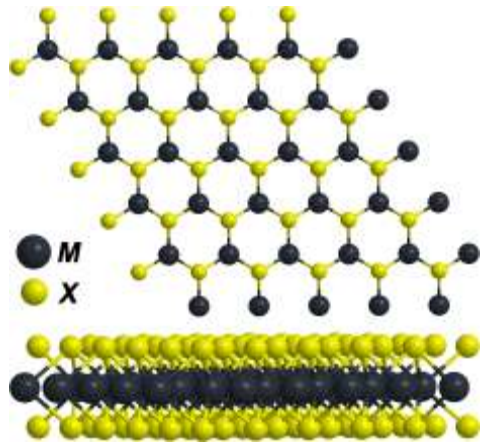

**Figure 11.** General structure of TMDs. The monolayers are formed with a transition metal (*M*) bound to a chalcogen atom (*X*) in a hexagonal arrangement with no inversion center. The lower structure is a side view showing a monolayer consisting of a transition metal sandwiched between two layers of chalcogen atoms.

TMD monolayers can be fabricated by both 'top-down' and 'bottom-up' approaches[185]. The reliable production of TMD monolayers is essential for exploiting their well-defined electronic and optical properties and it should be noted that the Se-based TMDs tend to be more unstable than the other Group VIA element counterparts. Using a top-down approach, mechanical exfoliation takes advantage of the fact that TMD monolayers are crystalline and are easily cleaved due to weak inter-layer coupling by Van der Waals forces relative to the covalent bonds between *M* and *X* atoms. However, we note that in general, it would be desirable to begin with a crystal of high melting point as the melting temperature decreases with decreasing the thickness of thin films. And although the use of adhesive tapes to peel away layers is facile, this is done under ambient conditions which could cause corrosion and decomposition. Nonetheless, this technique does produce samples of monolayer material, typically about 5-10 μm in diameter that are suitable for device components[186] used in fundamental research, lab-based validation, and prototyping. And although substantial impurities are not introduced during mechanical cleavage, the major disadvantage is a low-production throughput and the difficulty in controlling the chemical and structural purity of the parent crystal material.[187]

The facile production of an abundance of TMD monolayer flakes is necessary for applications requiring scalable device integration. The 'bottom-up' chemical approach allows wafer-scale production of TMD monolayers, but this is yet to reach the level of production seen for graphene (30000 m$^2$ of graphene films per year as of 2014),[188] even though TMDs were studied earlier[186]. Nevertheless, wafer-scale synthesis methods employing chemical vapour deposition (CVD) were used to obtain various TMD films.[189, 190] However, the control of the layer thickness of the TMD films over extended surfaces has not yet been unambiguously demonstrated.[190, 191]

Chemical exfoliation of TMD monolayers represents a high-yield approach.[185, 192] Briefly, Li$^+$ is electrochemically intercalated between the Van der Waals interlayers of the bulk TMD crystals. Exposing the intercalated Li$^+$ to water triggers a vigorous reaction between the layers whereby the evolved hydrogen gas rapidly expands and consequently separates the layers. The resulting aggregates may contain TMD material that are metallic. However, reproducibility is a major challenge for chemical exfoliation methods to TMDs: the trade-off involves the production of gram quantities of TMD monolayers, yet these quantities consist of an ensemble of aggregates differing structurally and electronically from the bulk material[193]. Molecular beam epitaxy (MBE) was also demonstrated to prepare high-quality epitaxial films on graphene-terminated 6H-SiC(0001) substrate.[194] MBE also allowed for the fine control of non-stoichiometric phase preparation.[195] There are many other methods to synthesise layered TMDs, for example ultrasonication in solvents[196], physical vapour deposition[197], hydrothermal synthesis[198], and more recently, Tedstone et al.[199] described a number of methods to synthesise TMD materials with transition-metal dopants. The versatility in the production of TMD materials could pose discussions around well-defined and reproducible physical properties.

Bulk TMDs are indirect gap semiconductors with 1-2 eV band gap and with a valence band maximum located at the Γ point and a conduction band minimum located at a low-symmetry point of the Brillouin zone (Figure 12)[37]. Increasing the size of the *M* atom leads to an increase in band gap while increasing the size of the *X* atom reduces the band gap in the *MX*$_2$ TMD structure. The shapes of valence and conduction bands, however, undergo significant changes upon decreasing the number of TMD layers. The position of the gap shifts to the K point, tending towards a direct gap semiconductor for monolayer TMDs. For instance, monolayer MoS$_2$ shows four orders of magnitude larger luminescence quantum efficiency compared to the bulk 3-dimensional material, as a consequence of the difference in the band gap change.[177] Also, angle-resolved photoemission spectroscopy provided a method to directly probe the dispersion of the valence band, confirming the predicted shift of the valence band maximum (VBM) from Γ to K upon reduction of the thickness of MoS$_2$ films to a single layer.[200]

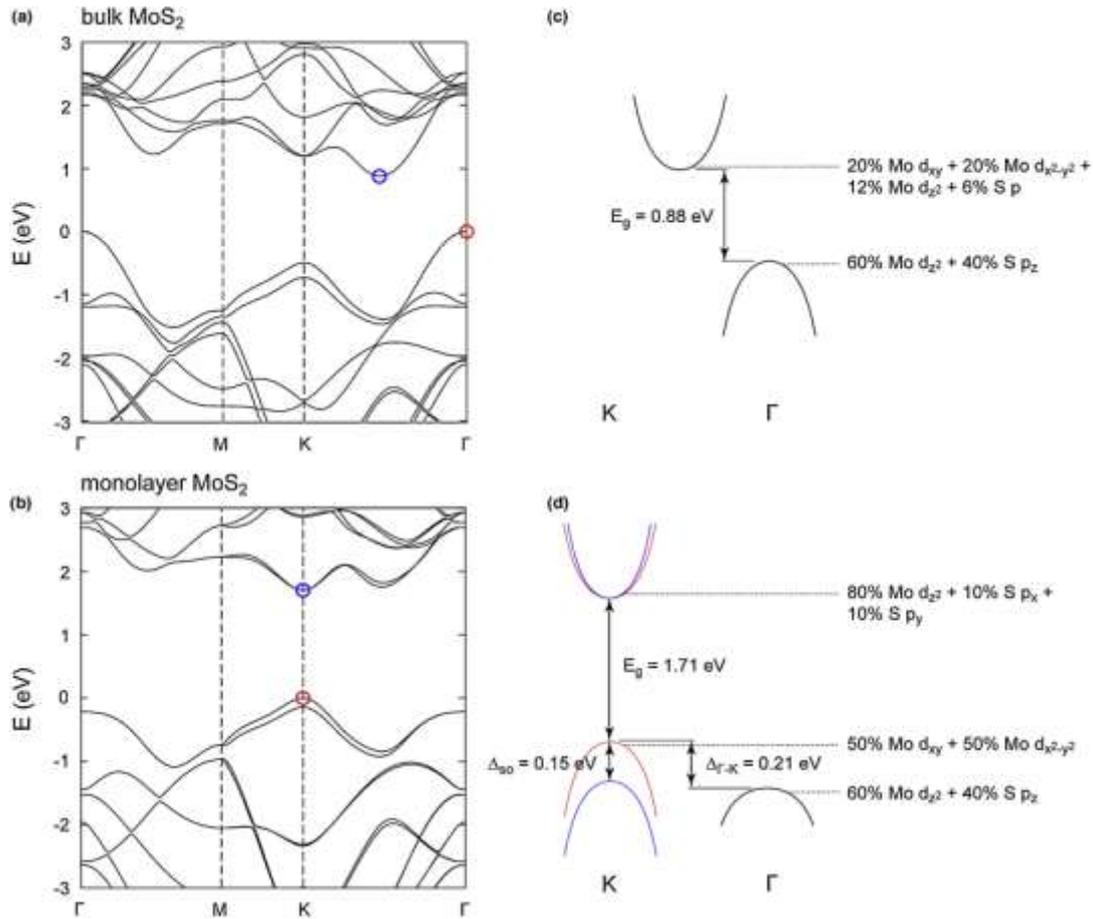

**Figure 12.** Electronic band structure of $MoS_2$. (a) Bulk $MoS_2$ showing an indirect band-gap and (b) monolayer $MoS_2$ showing a direct band-gap. Electronic bands calculated from first principles using density functional theory with the generalised gradient approximation. Valence band maxima and conduction band minima are indicated by red and blue circles, respectively. Energies are given relative to the valence band maxima. (c) bulk $MoS_2$ and (d) monolayer $MoS_2$ showing the band gaps $E_g$ as well as the valence band spin-orbit splitting $\Delta_{so}$ and the $\Gamma$ valley band offset $\Delta_{\Gamma-K}$ for the case of monolayer $MoS_2$. Reproduced from Ref. [57] under a Creative Commons license.

Pristine TMD monolayers are diamagnetic insulators: for use as an active component in spintronics this would be undesirable[7]. Doping could provide routes to tune the intrinsic properties of TMD monolayers and provide electron spin-based functionalities. One possibility would be to use TMDs as non-magnetic channel materials in 'traditional' spin-valve configurations where the current flow carries the spin signal and is detected by magnetoresistance. The requirement for such a non-magnetic channel would be to transport spin currents with minimal spin relaxation. This not-so farfetched proposal was tried with $MoS_2$ vertically sandwiched between Permalloy (an alloy of Ni and Fe that is easily magnetised and demagnetised) electrodes that demonstrated a spin-valve effect below 240 K. However, only a very small (0.37 %) maximal magnetoresistance was obtained (Figure 13).[201] This comparatively low value compared to the theoretically predicted 9 % maximal attainable magnetoresistance indicated that there was room for device optimisation due to the nominally insulating MoS$_2$ layer showing metallic behaviour (perhaps due to strong hybridisation with the Ni and Fe atoms at the MoS$_2$-Permalloy interface).[201] To increase magnetoresistance Tarawneha et al.[202] proposed the use of lateral heterojunction geometry (i.e. the current flowing in the plane of the MoS$_2$ monolayer). In this case, the authors used non-equilibrium Green's functions in the framework of a density functional approach, to calculate that a large 150% magnetoresistance in Fe/MoS$_2$/Fe hetero-junctions could be possible.[202]

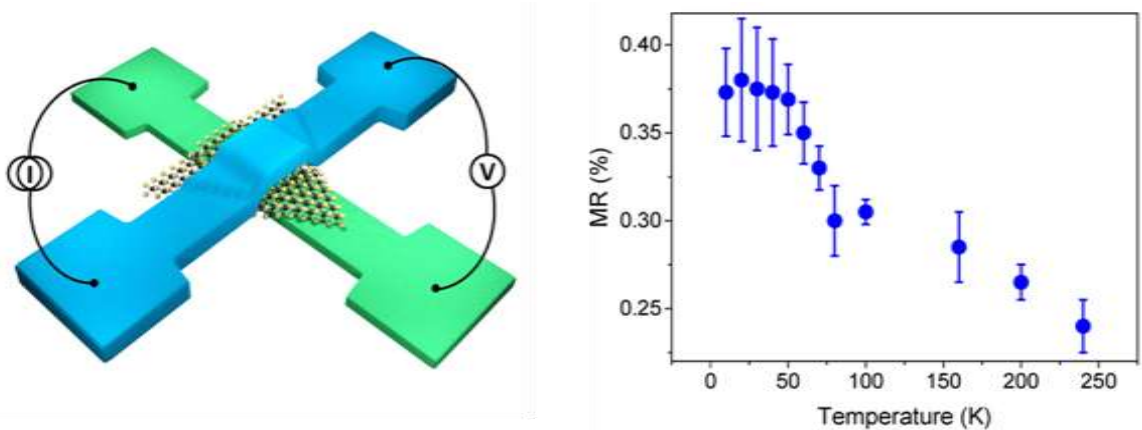

**Figure 13.** A schematic diagram of the four-terminal spin-valve device based on a MoS$_2$ monolayer. In this current perpendicular to plane geometry the current flows vertically through the junction area. The magnetic field is applied in-plane along to the bottom electrode. The magnitude of the magnetoresistance, MR, as a function of temperature is plotted on the right. Adopted with permission from Ref. [201]. Copyright 2016 American Chemical Society.

At 5 K, for both MoS$_2$ and WS$_2$, the spin lifetimes obtained were ~3 ns and would be expected to rapidly diminish at room temperature.[203, 204] This is because with increasing temperature, the lattice phonon modes together with the spin-orbit coupling in these non-magnetic systems containing heavy elements would be expected to dominate the spin dynamics[12]. Moreover, the inversion symmetry is absent in TMD monolayers, thus both Elliott-Yafet and D'yakonov-Perel' mechanisms play a role in the spin dynamics.[13, 14, 15, 205] The spin lifetime in MoS$_2$ and WS$_2$ were also found to be inversely proportional to the momentum relaxation time. Nevertheless, this spin-relaxation time is longer than the typical exciton recombination times of a few picoseconds[203]; a scenario exists then where coupled spin and valley dynamics could result because of the spin dephasing of itinerant electrons in the rapidly fluctuating spin-orbit field in TMDs.[203]

In monolayers of TMDs inversion symmetry breaking together with spin-orbit coupling can lead to the coupling of spin with energy degenerate valleys at the corners of the first Brillouin zone[206]. Selective photoexcitation of carriers with various combination of valley and spin indices can be achieved when the optical interband transitions have frequency-dependent polarisation selection rules[180, 207]. Kioseoglou et al.[181] used circularly polarised light to investigate the coupling of the valley and spin indices in monolayer MoS$_2$ and MoSe$_2$ to the depolarisation of emitted light. From this investigation they found that the origin of the spin relaxation processes involved phonon-assisted intervalley scattering and that the depolarisation and intervalley scattering were governed by the excess energy imparted to the photoexcited carriers through optical pumping. This explanation was extended to single-layer $WS_2$ films where the photoluminescence was from either the neutral or charged exciton[182]. However, the thermally activated relaxation of the carriers may not represent all aspects of physics describing the spin relaxation in TMDs[55].

Mak et al.[208] found that a lower bound hole valley-spin lifetime of 1 ns in monolayer $MoS_2$ was due to the intervalley scattering from the K to K' point. The intravalley hole spin relaxation from interlayer tunnelling of charge carriers could be described using the Elliot-Yafet mechanism as well as electron-hole exchange interactions[209]. Mak et al.[208] extended these explanations to describe the spin relaxation of a few hundreds of femtoseconds in bilayer $MoS_2$. This much shorter lifetime compared to monolayer $MoS_2$ was fundamentally due to the indirect bandgap in bilayer $MoS_2$. However, in bilayer $MoS_2$, even though a net spin orientation exists, there is no valley polarisation: bilayer $MoS_2$ is useless in valleytronics. The application of an electric field perpendicular to the bilayer may restore the spin-orbit splitting and thus render bilayer $MoS_2$ useful for valleytronics applications[210]. Interestingly, the generation of charged exciton states in layered TMDs opens the possibility of using gate voltages to modulate the polarisation (or intensity) emitted from the TMD structures[182].

As we have seen in the previous sections of this feature article, magnetisation control can be achieved via spin-orbit coupling. Strong current-induced spin-orbit torques induced at a monolayer $MoS_2$-Permalloy interface have been reported recently.[211] To effectively capitalise on this opportunity, it would be advantageous to impose magnetic control over the otherwise intrinsically nonmagnetic and insulating TMDs. This would result in the much sought-after dilute magnetic semiconductors[212]. A large body of computational predictions continue to search for suitable magnetic dopants on TMDs including V, Cr, Mn, Fe, Co and Cu, that would yield localised magnetic states and magnetically ordered phases to pave the way for a viable approach to spintronics using TMDs.[213, 214-216]

Among the large number of dopants investigated, there has been vigorous analyses of Fe on $MoS_2$; possibly because of the high total magnetic moment of Fe and the greater likelihood of producing an *n*-type ferromagnetic semiconductor[217]. Shu et al.[214] predicted that the number of layers was crucial in controlling the magnetic properties and that monolayer $MoS_2$ doped with Fe was predicted to be ferromagnetic, while bi- and tri-layer Fe-doped $MoS_2$ became antiferromagnetic. DFT calculations were also used to predict that mechanical strain also had a large impact on the magnetic properties of the Fe-$MoS_2$ system; the spin system reorients upon a small 3 % biaxial tensile strain.[215] We could reasonably expect an additional facet in the research area of magnetically doped TMDs to come from structural confinement effects. Recent reports have emerged of nanoribbons of $MoS_2$ in armchair configurations predicted to be nonmagnetic while the zigzag conformation being predicted to become metallic and magnetic with 3*d* transition metal doping.[216]

Copper doping of various TMDs has also been computationally studied recently. For $MoS_2$ and $MoSe_2$ even a single Cu dopant atom[218, 219] gave a surprisingly high total magnetic moment of about 5 $\mu_B$, higher than any of the other first-row transition metals.[45,46] The strong hybridisation of the 3*d* orbitals of Cu with the *p* orbitals of S atoms enabled spin splitting near the Fermi level and the consequent development of magnetic properties. On the contrary, in the case of $MoTe_2$ and $WS_2$ monolayers, Cu doping was not predicted to yield a magnetic material.[218] Ferromagnetism could be obtained in calculations for Cu-doped $WS_2$ bi-layers, however, with a total magnetic moment of about 1 $\mu_B$, significantly lower than that in the case of $MoS_2$.[220]

Experimental investigations of inducing magnetisation by intercalation are still lacking. Intercalating Mn into $MoS_2$ monolayers has been reported[221] with doping levels of ~2 at.% Mn, however this caused a loss of two-dimensionality due to competing MnS 3-dimensional structures. Nevertheless, weak ferromagnetism in $Sn_{1-x}Mn_xSe_2$, even at room temperature has been reported.[195] Cheng et al.[222] demonstrated that several transition metals can be incorporated to $WS_2$ nanoflakes, and while their doping with $Gd^{3+}$ was employed for magnetic resonance imaging, one could easily envision translating such materials into spintronic applications. The synthesis of Co-doped $MoS_2$ nanosheets can also be synthesised hydrothermal methods with cobalt atoms incorporated at edge sites.[223] The doped samples exhibited room-temperature ferromagnetism, however the authors noted that the total magnetic moment was observed to reduce due to morphological changes while varying Co concentration.[223]

The measured values of carrier spin lifetimes in TMDs at low temperatures are currently low (<3 ns) and at room temperature (<<1 ns) are impractical. The reported mobilities for TMDs are low. However, interest in TMDs is likely to continue growing, especially with recent reports of the synthesis of atomically thin and stable $ZrS_2$.[224] Due to the small effective mass in $ZrS_2$, the theoretical upper limit of acoustic-limited carrier mobility is about 4 times larger than in the cases of $MoS_2$. Additionally, due to the smaller band gap of 1.4 eV, the optical absorption edge lies in the visible spectral range, which may allow for practical optoelectronic applications. So although TMDs pose compromising aspects towards the much sought-after dilute magnetic semiconductors[212, 217], it might be possible that TMDs would more immediately find their use as valuable components in multifunctional layered heterostructures (not necessarily limited to spintronic devices), contributing to useful compositions beyond their own limited permutations[35, 45, 225]. The use of TMDs as a platform material in a number of potential devices necessarily depends on the working dimensions of TMDs being well-defined in order to reproduce the quantum characteristics of TMD-based devices: a move towards scaling TMD-based devices could be considered quite bold.

## 6. Covalent Heterostructures

In the previous sections of this feature article, 2-dimensional materials were considered comprised of a single- or few-atomic layers of a single chemical compound. Devices, however, are rarely comprised of a single pristine free-standing monolayer material. Substrates and capping layers are used for adding additional functionalities, like mechanical support, electronic gates, and robust isolation from chemical environments. These layers are usually fabricated in a manual step-wise manner to contribute towards an independent function. A conceptually different avenue to devices incorporating two-dimensional materials would be to take advantage of the principles of molecular self-assembly to design 3-dimensional crystals whereby 2-dimensional layers are stacked to give multi-functionality. In such a case, the interactions between layers would be strongly covalent in all 3 spatial directions giving rise to greater stability over heterostructures predominantly van der Waals in nature[35], and there would exist an electronic distinction between the 2-dimensional layers as a result of the constituent layers exhibiting strong anisotropy.

Such materials have been found to exist in the form of layered π-functional molecular crystals.[46] Prime examples are TTF-derivative based molecular metals where conducting layers may form, for example, in the compound bis(ethylenedithio)tetrathiafulvalene (BEDT-TTF). The electronic bands of the quasi-two-dimensional π-functional molecular conductors are derived from molecular orbitals. The principles behind their use in the construction of crystalline organic 'metals' are conceptually very simple: free charges conduct through π-like molecular orbitals of neighbouring open shell molecules rather than between atomic orbitals in regular metallic conductors.[226] For example, as individual BEDT-TTF molecules are surrounded by voluminous molecular orbitals, in order to create continuous electronic bands, it would merely be necessary to stack the BEDT-TTF molecules close enough so that the molecular orbitals can overlap. This would enable electron transfer from one molecule to another.

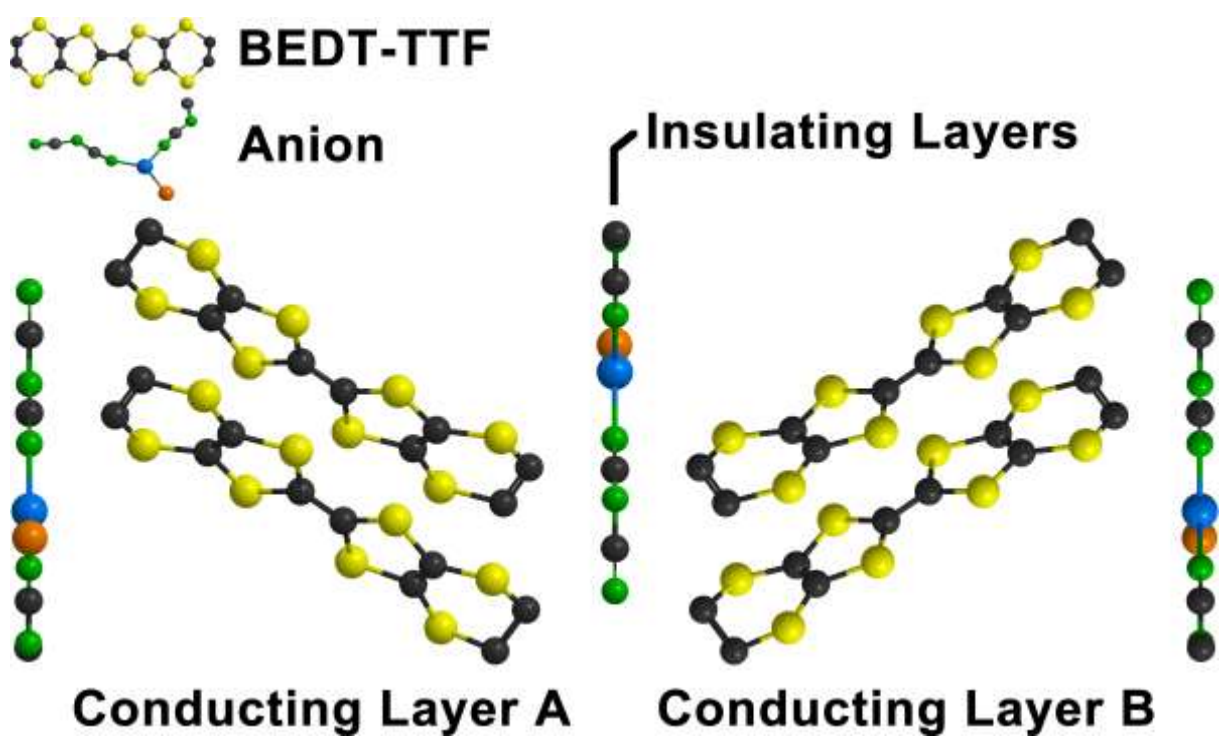

**Figure 14**. Schematic crystal structure of the κ-(BEDT-TTF)$_2$Cu[N(CN)$_2$]Br charge transfer salt. For clarity one pair of BEDT-TTF molecules is shown per layer and hydrogen atoms are omitted. The structure of the BEDT-TTF cation and the Cu[N(CN)$_2$]Br anion are shown.[227]

The conduction through an overlapping of band structures formed by the ordered self-assembly of π-functional molecular crystals could then be accomplished by formulation of a charge-transfer salt.[46, 226, 228] In a charge-transfer salt (and still using BEDT-TTF as an example), a number (*N*) of BEDT-TTF molecules will jointly donate an electron to a second type of molecule, the anion (Λ), to form the compound α-(BEDT-TTF)$_N$Λ. The transfer of charge serves to bind the charge-transfer salt together in a manner analogous to ionic bonding, and also leaves behind a hole, jointly shared between *N* BEDT-TTF molecules. Hence, the bands formed by the overlap of the BEDT-TTF molecular orbitals will be partially filled, leading to electrical conductivity. An example of an anion is Cu[N(CN)$_2$]Br$^-$ polymer, which forms a charge-transfer salt κ-(BEDT-TTF)$_2$Cu[N(CN)$_2$]Br (Figure 14), a $T_C$=12 K superconductor[227]. Most of the molecules used to create quasi 2-dimensional organic conductors are of a similar form to BEDT-TTF. For example, when the innermost four sulphur atoms of BEDT-TTF are replaced by Se, one obtains BETS-TTF[229], and other varieties include asymmetric EDT-TTF[230], DMET[231] (dimethyl(ethylenedithio)diselenadithiafulvalene)), and coronene[232].

It is important to note that the electronic interactions are not only determined by the distance between adjacent molecules but also by their spatial arrangement. The α-type structure of BEDT-TTF consists of stacks arranged in a herring-bone pattern, whereas the κ-type phase consists of two face-to-face aligned molecules with the adjacent molecules arranged almost orthogonal to each other. It is the κ-type phase arrangement that gives rise to the quasi 2-dimensional electronic structure and anisotropy. The BEDT-TTF molecules are in close proximity to each other within the layers, allowing substantial overlap of the molecular orbitals; the transfer integrals[226], which parameterise the ease of hopping of electrons between BEDT-TTF molecules, would then be relatively large within the BEDT-TTF planes. Conversely, in the direction perpendicular to the BEDT-TTF planes, the BEDT-TTF molecules are well separated from each other; the transfer integrals would be much smaller in this direction (i.e. hopping is more difficult) resulting in extreme anisotropy. For example, the electronic anisotropy of κ-(BEDT-TTF)$_2$Cu[N(CN)$_2$]Br was measured[47] to be $\rho_\perp/\rho_\parallel$~10$^5$-10$^6$. This meant that a single unit cell thick material of 3 nm would be a complete field-effect transistor with μ~10 cm$^2$V$^{-1}$s$^{-1}$ mobility[233] and a gate electrode. Indeed research in this direction has demonstrated field-effect devices with ionic liquid gating.[234] These devices also demonstrated unique properties like gate- or light-induced control on superconductivity at low-temperatures partially due to the strongly correlated nature of the electron, which is absent in van der Waals heterostructure systems. The electronic anisotropy in layered covalent heterostructures provide opportunities to develop both novel electronic and spintronics devices.

The magnetic properties in layered 2-dimensional π-functional molecular crystals originate from the conducting π-electron system that gives rise to Pauli paramagnetism. This, combined with the remarkably long spin diffusion length of 0.2 µm at room temperature, makes these materials suitable candidates as spintronic wires.[47] The long spin-diffusion length[47] of κ-(BEDT-TTF)$_2$Cu[N(CN)$_2$]Br was also showed to be advantageous in recent spin-current injection experiments at room temperature.[235] A method for the direct conversion of a spin current into an electric signal, dubbed the 'inverse spin Hall effect', was utilised in the work of Qiu et al.[235], which has opened exciting new possibilities for the application of 2-dimensional organic materials in spintronic devices as spin-charge converters.

Other key properties of the κ-(BEDT-TTF)$_2$Λ material family in terms of applicability in the field of spintronics is that magnetism occurs in a homogenous fashion. This could be achieved by incorporating magnetic ions with localised moments.[236] Moreover, replacing the single Br atom in the unit cell with Cl has been shown to change the superconducting ground-state to a weak ferromagnetic one.[237] Furthermore, since the constituent elements are of light atomic weight, the spin-orbit coupling is weak and the spin lifetime and spin diffusion length could provide workable values.[47, 48]

To achieve the same spin valve functionality, the use of a single unit cell thick κ-(BEDT-TTF)$_2$Cu[N(CN)$_2$]Cl was proposed.[48] The principle behind this proposition was that in order to allow for the separate control of the magnetisation of the layers in a spin valve one layer is usually made "hard" by pinning its magnetisation to another (third) antiferromagnetic layer. And in this regard, it was shown that the magnetic order of the adjacent BEDT-TTF layers were magnetically independent. Magnetic coupling was as small as 12 neV according to ESR measurements.[238] In this case, the long-range magnetic order and thus the magnetisation direction was set by in-plane magnetic anisotropies of each BEDT-TTF layers according to the Mermin-Wagner theorem.[239] Modifying the in-plane anisotropy thus resulted in a strong change of magnetic order and gave rise to the spin valve effect. And the *in-situ* modification of anisotropy for κ-(BEDT-TTF)$_2$Cu[N(CN)$_2$]Cl was demonstrated by application of external magnetic field.[48, 240] Since the dominant anisotropy of the planes was the Dzyaloshinskii-Moriya interaction, individual layers could be addressed separately by appropriately orienting the magnetic field.[48] Indeed, the zero-field Néel temperature $T_N$=23 K was shifted to 32 K by 8 T field, which was a remarkable 50% increase.[48]

The observation of multiferroicity was made by Lunkenheimer at al.[241] in the κ-(BEDT-TTF)$_2$Cu[N(CN)$_2$]Cl system. In the context of multiferroics – where long-range magnetic and electronic dipole order coexists – this represented a significant step towards the tantalising thought of future data storage and processing devices based on organic materials. Especially, when considering that in multiferroics magnetism can be controlled by an electric field rather than an electric current to significantly reducing device energy consumption, and there exist opportunities to integrate spintronic functionality.[242] To add to such a paradigm shift, it was also demonstrated that the spin order in κ-(BEDT-TTF)$_2$Cu[N(CN)$_2$]Cl was driven by ferroelectricity, in marked contrast to the spin-driven ferroelectricity in non-collinear magnets.[241, 243] Thus, κ-(BEDT-TTF)$_2$Cu[N(CN)$_2$]Cl not only represents a new class of multiferroics, but would also be seriously considered as a candidate for the realisation of a new coupling mechanism of magnetic and ferroelectric ordering.

The carriers in covalent heterostructures originate from the π-electron system which are itinerant and currently show room temperature lifetimes (<5 ns) below practical values of 100 ns. The carrier mobilities obtained so far are low. However, the spin diffusion lengths in these types of molecular crystals are long. Together these intrinsic properties pave the way for new types of spintronics devices based on material components functioning as 'spin-wires'. A great advantage in such two-dimensional organic materials is the versatility in chemical control of morphology and constituent atoms that would allow device components to be engineered ad-hoc: a similar reason for the successful widespread use and scalability of metal-organic frameworks. This chemical versatility demands reproducibility in the purity and stability of the structure. Beyond systematic chemical synthesis and characterisation, a more concerted and targeted approach would be able to identify suitable compounds for detailed spintronics investigations. The κ-(BEDT-TTF)$_2$Λ material family are gaining traction in the field of spin-electronics and are shedding light on what may prove to be an important, extendable class of 2-dimensional layered materials.

## 7. Topological Materials

Topological phases are insensitive to smooth changes in material parameters and cannot change unless the system passes through a quantum phase transition[49]. The spin-orbit interaction can lead to topological insulating electronic phases[109, 244]. Topological insulators can be understood within the framework of the band theory of solids[245]. The two-dimensional topological insulator is known as a quantum spin Hall insulator, that exists at zero external magnetic field[51, 244]. A topological insulator, like an ordinary insulator, has a bulk energy gap separating the highest occupied electronic band from the lowest empty band. The surface of a topological insulator (referred to as an edge in the two-dimensional case), however, necessarily has gapless states that are protected by time-reversal symmetry[246]. The metallic edge states of a topological insulator have a distinct helical property: spin-polarised two-dimensional Dirac fermions counter propagate at a given edge[109, 247].

The first topological insulators were experimentally realised in HgTe quantum wells sandwiched between CdTe layers[51]. These HgTe/(Hg,Cd)Te quantum well structures were fabricated with tunable carrier densities and mobilities that allowed for the observation of a quantum phase transition that provided evidence of the quantum spin Hall effect. The strongly anisotropic nature of the magnetoresistance was demonstrated by applying a small magnetic field perpendicular to the two-dimensional electron gas plane destroying the quantum spin Hall effect. In this case, a magnetic field broke the time-reversal symmetry and thus resulted in an energy gap between the two otherwise degenerate helical edge states. It was found that this strong anisotropy originated from the high Fermi velocity of the edge states and the small bulk energy gap, which together made the orbital magnetisation dominant.

Soon after the observation of the topologically protected states in HgTe/(Hg,Cd)Te, several compounds were theoretically predicted to be three-dimensional topological insulators[248].

Amongst these compounds, $Bi_2Se_3$ and $Bi_2Te_3$ crystals, of general tetradymite-type $Bi_{1-x}Sb_x$, were experimentally identified to be large-gap topological insulators with a single Dirac cone on the surface (Figure 15)[249, 250]. Furthermore, spin resolved photoemission spectroscopy (PES) experiments have supported the theoretical prediction that backscattering is forbidden in these crystals due to time reversal symmetry, originating from the topological nature of the surface states.[251]

The crystal structure of $Bi_2Se_3$ contains a quintuple layer with the Se-Bi-Se-Bi-Se sequence (Figure 15a). The chemical bonds within a quintuple layer are strong, however, adjacent quintuple layers are only weakly bound by Van der Waals interactions. Mechanical separation of the quintuple layers is also achievable[252]. Moreover, in the Van der Waals gap of the quintuple layers various magnetic dopant ions like Cu, Mn, and Fe have been intercalated.[253] This structural and chemical flexibility together with the large bulk gap and the simplicity of the surface-state makes these Bi-based topological insulators attractive candidates for spintronics applications. However, spin dynamics studies on these more recent second-generation Bi-based topological insulators remain relatively unexplored.

PES is often employed to directly probe topological surface states and surface-bulk electronic correspondence[254]. These static experiments can be extended to include time-resolved studies of the dynamics of the photo excited states[255, 256]. Scanning tunnelling spectroscopy (STS) can also be employed in the study of topological surface states[257]. However, similarly to PES, STS lacks access to spin dynamics. The combination of pumped probe and spin resolved PES could give new insights to spin dynamics in topological materials, but these measurements are currently unfeasible due to the limited radiant fluence of pumped sources. Although, recent pumped probe PES experiments resolved remarkably long charge lifetime of over 4 μs in $Bi_2Te_2Se$.[256] This charge lifetime value according to the Elliott-Yafet model presents a lower bound for the spin-lattice relaxation time, $T_1$.

Mn-doped $Bi_2Se_3$ thin films show magnetically induced spin reorientation simultaneously with a Dirac-metal to gapped-insulator transition[253]. The electronic ground state of the Mn-doped $Bi_2Se_3$ exhibits unique hedgehog-like spin textures, which directly demonstrate time reversal symmetry breaking on the surface. The effects of Mn-doping were accompanied by the disappearance of the topologically protected surface state which manifested as an insulating gap.

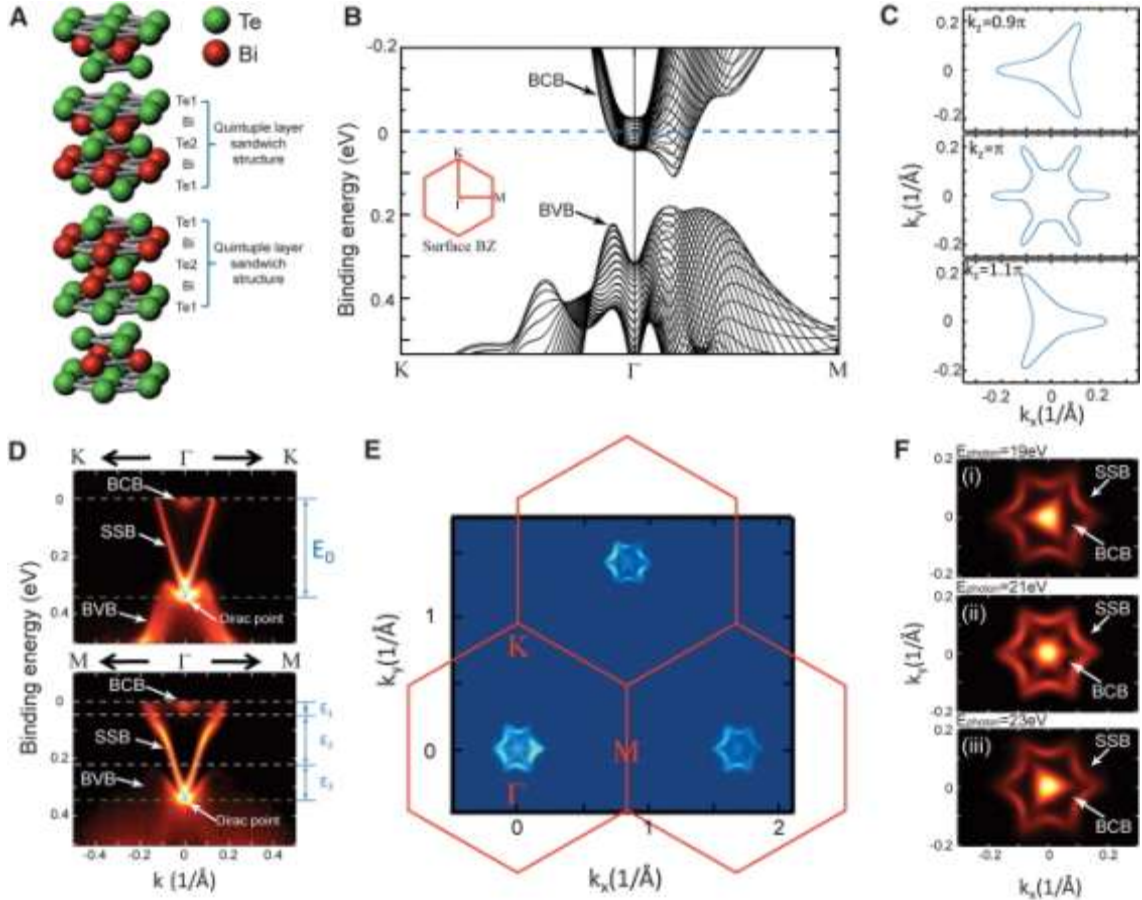

**Figure 15.** Crystal and electronic structures of $Bi_2Te_3$. (a) Tetradymite-type crystal structure of $Bi_2Te_3$. (b) Calculated bulk conduction band (BCB) and bulk valance band (BVB) dispersions along high-symmetry directions of the surface Brillouin zone (BZ), with the chemical potential rigidly shifted to 45 meV above the BCB bottom at Γ to match the experimental result. (c) The $k_z$ dependence of the calculated bulk Fermi surface projection on the surface BZ. (d) Angle resolved PES measurements of band dispersions along K-Γ-K (top) and M-Γ-M (bottom) directions. The broad bulk band (BCB and BVB) dispersions are similar to those in (b), whereas the sharp V-shape dispersion is from the surface state band (SSB). The apex of the V-shape dispersion is the Dirac point. Energy scales of the band structure are labelled as follows: $E_0$: binding energy of Dirac point (0.34 eV); $E_1$: BCB bottom binding energy (0.045 eV); $E_2$: bulk energy gap (0.165 eV); and $E_3$: energy separation between BVB top and Dirac point (0.13 eV). (e) Measured wide-range Fermi surface map covering three BZs, where the red hexagons represent the surface BZ. The uneven intensity of the Fermi surfaces at different BZs results from the matrix element effect. (f) Photon energy-dependent Fermi surface maps. The shape of the inner Fermi surface changes markedly with photon energies, indicating a strong $k_z$ dependence due to its bulk nature as predicted in (c), whereas the non-varying shape of the outer hexagram Fermi surface confirms its surface state origin. From reference[250]. Reprinted with permission from AAAS.

The remarkable robustness of the topologically protected surface states against adatom doping was demonstrated by the PES technique on the $Bi_2(Be,Se)_3$ family of topological insulators[258]. The topological insulating state in $SmB_6$ has been found to survive virtually any surface treatment by ion bombardment[259]. The robustness together with the spin polarisation and suppressed backscattering make two-dimensional topological surface states an attractive platform for high mobility charge- and spin-transport devices. However, controlling the stoichiometry of materials to display topological surface states remains difficult. The nominally insulating bulk is often heavily doped leading a very large contribution of the bulk states to the transport properties making it difficult to deconvolute the surface contributions. At low temperatures, however, the helical state can be detected by means of electronic transport measurements[260]. At close to room temperatures in materials where the surface states dominate the conductivity, as in the highly insulating $BiSbTeSe_2$, the surface-bulk correspondence could be separated and the observation of well-developed half-integer quantum Hall effect arising from topological surface state was reported.[50] One of the biggest challenges that remains in the development of devices incorporating topological insulator components has been in overcoming the inadvertent masking of the electron charge and spin properties by the existence of the bulk.

Giant and linear magnetoresistance was measured in $Bi_2Te_3$ nanosheets with a thickness below 20 nm[261]. Remarkably, this magnetoresistance exceeded 600% at room temperature. This value increased towards higher temperatures, showed a weak temperature dependence, and a linearity with field without any sign of saturation at measured fields up to 13 T. The linearity of the magnetoresistance was accounted for by the quantum magnetoresistance model proposed for zero-gap band structures with Dirac linear dispersion[262]. This observed linear giant magnetoresistance could pave the way for topological insulators to be considered for practical applications in magnetoelectronic sensors such as disk reading heads.

The quantum description of spin-1/2 particles is given by the solutions of the Dirac equation[263]. The Dirac equation can be split into a system of two equations whose solutions are distinguished by chirality. Weyl fermions appear when two electronic bands cross[264]. The crossing point is called the Weyl node, away from which the bands disperse linearly in momentum space. The surface of a material which contains Weyl fermions could then exhibit a new type of surface state: an open Fermi arc that would connect two Weyl nodes and then continue on the opposite surface of the material. Each Weyl point is chiral and contains half the degrees of freedom of a Dirac point, and can be viewed as a magnetic monopole in momentum space.

The existence of Weyl fermions as quasi particles in TaAs, TaP, NbP and NbAs was predicted[265] and has been recently followed by predictions of many more Weyl semimetal compounds[184, 266]. Weyl semimetals broaden the classification of topological phases of matter beyond insulators. A Weyl semimetal is a gapless metal which hosts Weyl fermions and has a topological classification that protects Fermi arc surface states on the boundary of a bulk sample[52, 267]. A band structure like the Fermi arc surface states would violate basic band theory in an isolated two-dimensional system and can only exist on the boundary of a three-dimensional sample. The existence of such isolated systems provide another example of the surface-bulk correspondence in a topological phase. The topological character of Weyl semimetals suggests that their exotic properties could be robust against decoherence, thus making them ideal candidates for applications in spintronics and quantum computing. And unlike topological insulators where only the surface states are of interest[49], a Weyl semimetal features unusual band structures in the bulk and on the surface[268].

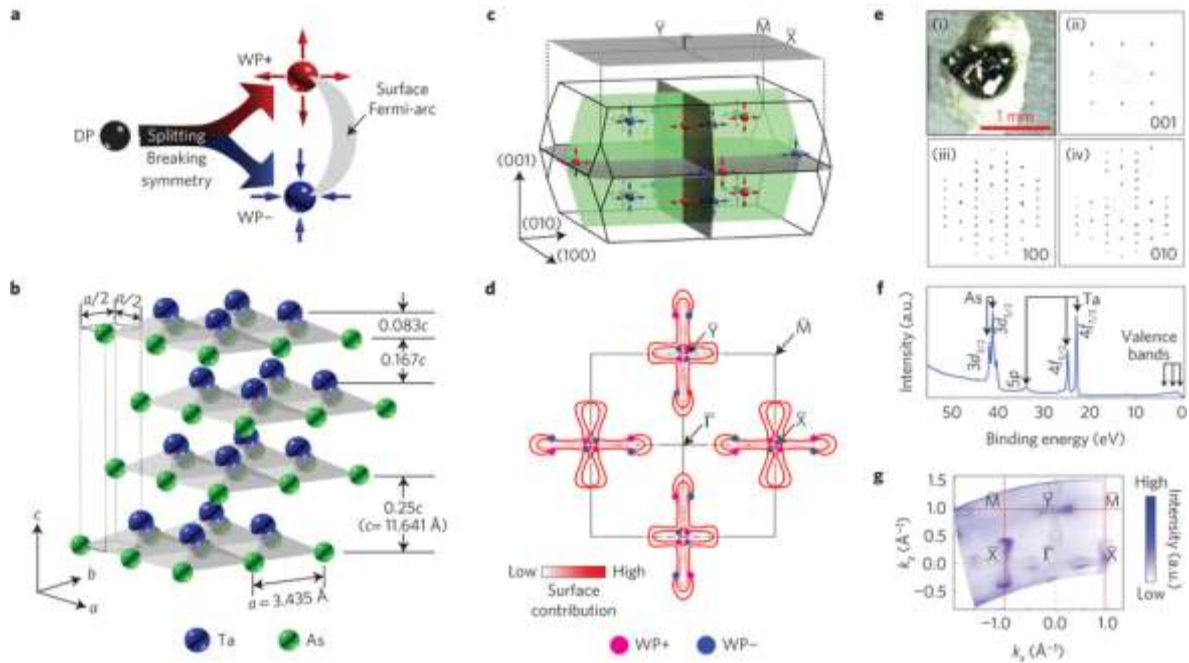

**Figure 16.** (a) Illustration of the splitting of a Dirac point (DP). A DP can be split into a pair of Weyl points with opposite chirality (marked as WP+ and WP-) which behave as a 'source' or 'sink' of Berry curvature by breaking time-reversal or inversion symmetry. The two Weyl points are connected by the Fermi-arc-type of Fermi surfaces formed by the topological surface states. (b) Crystal structure of TaAs, showing the A–B–C–D stacking of TaAs layers. (c) Schematic of the bulk and (001) surface Brillouin zones (BZs) of TaAs. Twelve pairs of Weyl points are predicted in each BZ, with four pairs at each of the $k_z = 0$ and $\pm 1.16\pi/c$ planes, respectively. (d) Fermi surfaces from *ab initio* calculations are plotted on the (001) surface BZ with the (projected) Weyl points (in red and blue) overlaid, showing the characteristic Fermi-arc Fermi surface geometry. The colour bar shows the surface contribution of the FS (white/0% to red/100%). (e) (i) Image of a TaAs single crystal with a flat cleavage plane used for angle resolved PES measurements. (ii–iv) X-ray diffraction patterns of the TaAs crystal from different crystalline directions. (f) Core-level photoemission spectrum clearly showing the characteristic As $3d$, Ta $5p$ and $4f$ peaks. (g) Broad Fermi surface map confirming the (001) cleavage plane and the lattice constant in (b). The uneven intensity of the Fermi surface at different BZs results from the matrix element effect. The colour bar shows the angle resolved PES spectra intensity, from white (lowest) to blue (highest). Reprinted by permission from Macmillan Publishers Ltd: Nature Physics (reference [269]), copyright 2015.

In 2015, Xu et al.[52] used PES to directly observe Fermi arcs on the surface of TaAs single crystals, as well as the Weyl fermion cones and Weyl nodes in the bulk, confirming TaAs as a Weyl semimetal. Yang et al.[269] also performed PES on TaAs to report its complete band structure, including the Fermi-arc Fermi surface and linear bulk band dispersion across the Weyl points (Figure 16). Lv et al.[53, 270] employed similar PES studies to directly observe Weyl nodes in TaAs. The spin texture of the Fermi arcs in TaAs have been also been detected by spin resolved PES demonstrating that the Fermi arcs are spin polarised[271]. The appearance of Weyl points near the Fermi level in TaAs causes novel transport phenomena related to the chiral anomaly, giving rise to negative magnetoresistance under parallel applied electric and magnetic fields[272]. These works have mounted strong evidence for the existence of Weyl fermions.

Weyl semimetals other than TaAs have been discovered. Xu et al.[54] used PES to discover the Weyl semimetal state in an inversion-symmetry-breaking single-crystalline solid NbAs. Wang et al.[273] experimentally demonstrated that the NbP Weyl semimetal had an unprecedented helical Weyl fermion charge carrier mobility of $10^7$ cm$^2$V$^{-1}$ at 1.5 K. Doping of various magnetic ions like Co, Mn, Gd into the structure was also demonstrated using binary transition-metal arsenides[274]. However, only 1% magnetic impurity doping by Cr caused the mobility of Weyl fermions in NbP to decrease by more than two orders of magnitude. This drop in mobility was attributed to the cancellation of helicity protection by magnetic impurities. This helicity protected Weyl fermion transport also manifested in the chiral anomaly induced negative magnetoresistance. In the ZrTe$_5$ three-dimensional Dirac semimetal, angular-dependent magnetoresistance measurements under high magnetic fields of up to 31 T revealed negative longitudinal magnetoresistance induced by the chiral anomaly[275]. It was proposed that these anomalies indicate the Dirac point splits into Weyl points due to broken time-reversal symmetry at high magnetic field.

Topological materials have shown exceptionally long and practical carrier lifetimes on the order of micro seconds at room temperature and exceptional low temperature carrier mobilities. The surface metallic state is robust and endows topological materials with readily diffuse spins carriers. However, the surface-bulk correspondence of 'large' single-crystals remains a significant challenge in engineering efficient device design: the bulk is essentially useless. However, topological protected states have been shown to exist in materials on the order of desirable device dimensions (thin films and particles of <20 nm thickness). The field of topological materials is now expanding at a rapid pace with the discovery of Weyl semimetals. And although some topological states like those in HgTe/(Hg,Cd)Te can be viewed as a tunable graphene system, where the Dirac mass term can be tuned continuously to zero from either the positive (topologically trivial) or the negative (topologically nontrivial) side, the topological stability of an isolated Weyl fermion is unlike two-dimensional massless Dirac fermions in graphene, where inversion symmetry of the honeycomb lattice is essential for their stability. This robustness has led to continued experimental and theoretical efforts to synthesising and optimising topological materials, characterising topological states by surface sensitive spectroscopy, transport measurements, device fabrication, and an extensive search for new material candidates.

## 8. Perspectives and Future Prospects

There are increasing motivations to explore new avenues to follow the ever-growing need for computational speed and storage capability. On the one hand, to cope with Moore's law, the existing electron-charge based semiconductor technologies are being pushed towards their limits in terms of both miniaturisation and enhanced performance, albeit, with still more room to manoeuvre[7]. On the other hand, to overcome speed and energy consumption limitations, other technologies that include spintronics, continue to be explored which are making use of an ever more coherent and robust quantum regime to address these limitations[91]. The most effective solutions offered by 2-dimensional materials to increase computational power may lie in more pragmatic approaches: an incremental addition to current computing architectures rather than a stand-alone and extremely disruptive technology[31]. Nonetheless, the research field of spin in nanomaterials has found an important and *current* problem to address[276].

Pushing the boundaries of workable size limits in the fabrication of nanotechnologies could be possible by utilising two-dimensional semiconducting and conducting materials like graphene, silicene, phosphorene, TMDs, numerous layered heterostructures and topological materials[277]. Graphene, because of its zero-band gap and the exceptionally high mobility, offers the opportunity to electronically connect multiple devices – ideally required for scalable chip fabrication. The other elementals, silicene and phosphorene, together with the TMDs, add to the growing library of materials that largely compensate for graphene's shortfalls, namely the zero-band gap, mediocre on-off ratios, and impractical quantum spin Hall effect. And despite the claims of contentious 2-dimenional materials of optical control, long-range magnetic ordering, silicon integration, and tunable band gaps, graphene currently remains a benchmark for many spin-electronic phenomena. However, new opportunities are emerging to literally displace graphene, which include using Van der Waals and covalent heterostructuring that offers labile and controlled functionality, and topological states of matter that are intrinsically robust against decoherence.

Although to date, the advent of room-temperature spintronic-based logic has ultimately been hindered by the difficulty to achieve both long spin lifetimes and coherent spin control simultaneously. Spin control is usually achieved by sizeable spin-orbit coupling. However, large spin-orbit coupling tends to lead to fast spin decoherence. This is a major bottleneck, which demands a continued and concerted effort between experiment and theory. It is partly in this regard that the mature field of graphene exhibits some superiority when contending for efficient spin information transmission and coherent spin manipulation, and partly due to its exceptional mobility values, the weak spin-orbit coupling in carbon, and good electrical conduction. The other atomic single layer semiconductors like phosphorene and silicene should not be completely discounted, as they could offer additional spin functionalities. The stronger spin-orbit interaction is sizeable in these elements, and cross coupling exists between electron momentum and the spin degree of freedom opening up the possibility for spin-valley electronics. Together with a finite and tunable band-gap this could offer direct electrical control for spin manipulation in these materials.[278] The TMDs provide an opportunity to confidently introduce long range magnetic order to the 2-dimensional world[199], which could expedite research efforts in this direction. Strong correlation effects in π-functional molecular crystals and the microscopic long-range quantum entangled nature of topologically protected surface states add additional flavours to 2-dimensional spintronics with the stabilisation of exotic spin states.

The combination of all these 2-dimensional material properties have ensured a lively and concerted research atmosphere into spin-based materials science and technology that is continually expanding. This, in large part, can be attributed to the need for 'end-to-end' solutions spanning a number of scientific disciplines and a depth in technological expertise. Looking back on more than a decade of research into 2-dimensional materials, seamless progress towards technology readiness would sensibly involve a multidisciplinary approach. To accelerate this process, the development of a targeted and systematic experimental toolkit would be needed. These instruments would provide the means to disentangle the contribution of spin lifetime from the intrinsic spin relaxation and extrinsic effects induced by disorder, interactions with substrate, adsorbed molecules etc. This is a formidable experimental challenge mainly due to the small volume of the studied materials and the conditions under which the materials are prepared and subsequently studied. Local or spectroscopic tools like scanning probe microscopes tipped with magnetic sensitivity, muon spin resonance, and ESR are crucial for characterising phenomena in 2-dimensional materials related to spintronic applicability. Furthermore, the qualitative aspects that indirectly impact spin dynamics like material reproducibility, compatibility, and improvements in fabrication techniques should not be judged any less important than the hard numbers crunched by measurement, as any industrial application will require a scalable approach. As such, there is currently no 'champion' 2-dimensional spintronic material: the most successful experiments to date have *simply* involved innovative materials construction.